\documentclass[12pt,epsf,amstex]{article}

\usepackage [dvips]{graphicx}
\usepackage{amsmath}
\usepackage{amssymb}
\usepackage{epsfig}
\usepackage{verbatim}
\usepackage{cite}

\addtocounter{secnumdepth}{1}
\usepackage[utf8]{inputenc}   
\usepackage[T1]{fontenc}
\usepackage{ae,aecompl}

\newcommand{\un}{ ${\mathbb{R}} \cup \ii{\mathbb{R}}$ }

\newcommand{\calT}{{\mathcal{T}}}

\newcommand{\calTzero}{$\mathcal{T}(0)$}
\newcommand{\calHzero}{$\mathcal{H}(0)$}
\newcommand{\calTone}{$\mathcal{T}(1)$}

\newcommand{\calHone}{$\mathcal{H}(1)$}
\newcommand{\calTn}{$\mathcal{T}(n)$}
\newcommand{\calHn}{$\mathcal{H}(n)$}
\newcommand{\calTnpone}{$\mathcal{T}(n+1)$}

\newcommand{\TtoH}{$\mathcal{T}\mapsto\mathcal{H}$}
\newcommand{\HtoT}{$\mathcal{H}\mapsto\mathcal{T}$}

\newcommand{\EGup}{{\rm EG}$_{\rm up}$}
\newcommand{\EGdown}{{\rm EG}$_{\rm down}$}
\newcommand{\EG}{{\rm EG}}
\newcommand{\ER}{{\rm ER}}
\newcommand{\parEGup}{{\rm (EG$_{\rm up}$)}}
\newcommand{\parEGdown}{{\rm (EG$_{\rm down}$)}}
\newcommand{\parEG}{{\rm (EG)}}
\newcommand{\parER}{{\rm (ER)}}

\newcommand{\vecr}{\vec{r}}

\newcommand{\ee}{\mbox{e}}
\newcommand{\ii}{{\rm i}}

\newcommand{\beq}{\begin{equation}}
\newcommand{\eeq}{\end{equation}}
\newcommand{\bea}{\begin{eqnarray}}
\newcommand{\eea}{\end{eqnarray}}

\numberwithin{equation}{section}

\newcounter{exerc}
\newcommand{\stepex}{\stepcounter{exerc}}

\newcounter{refa}
\newcounter{refb}
\newcounter{refc}
\newcounter{refe}
\newcounter{reff}
\newcounter{refg}
\newcounter{refh}
\newcounter{refi}
\newcounter{refj}
\newcounter{refk}
\newcounter{refl}

\newcounter{refp}
\newcounter{refq}
\newcounter{refr}
\newcounter{refs}

\newcounter{refu}
\newcounter{refv}

\begin{document}

\thispagestyle{empty}
\title{Iterated star-triangle transformation\\[2mm]
on inhomogeneous 2D Ising lattices} 
 
\author{H.J. Hilhorst\\[3mm]
Laboratoire Ir\`ene Joliot-Curie, B\^atiment 210\\
Universit\'e Paris-Saclay, 91405 Orsay Cedex, France\\}


\maketitle
\vspace{-1cm}
\begin{small}
\begin{abstract}

We consider infinite or periodic 2D triangular Ising lattices 
with arbitrary positive or negative nearest-neighbor couplings $K_i(\vecr)$,
where $\vecr$ and $i$ indicate the bond position and orientation, 
respectively. 
Iterative application of the star-triangle transformation 
to an initial lattice \calTzero\ with a set of couplings 
$\{K_i^{(0)}(\vecr)\}$ generates a sequence of lattices 
\calTn, for $n=1,2,\ldots,$ with couplings $\{K_i^{(n)}(\vecr)\}$.
When \calTzero\ includes sufficiently strongly frustrated plaquettes,
complex couplings will appear.
We show that, nevertheless, 
the variables $1/\sinh 2K_i^{(n)}\!(\vecr)$ 
remain confined to the union \un of the real and the imaginary axis. 
The same holds for a lattice with free boundaries,
provided we distinguish between ``receding''
and ``advancing'' boundaries,  
the latter having degrees of freedom that must be fixed by 
an appropriately chosen protocol. 
This study establishes a framework for future analytic and numerical work
on such frustrated Ising lattices.

\end{abstract}
\end{small}


\newpage


\section{Introduction} 
\label{sec_introduction}

\begin{figure}[ht]
\begin{center}
\scalebox{.40}
{\includegraphics{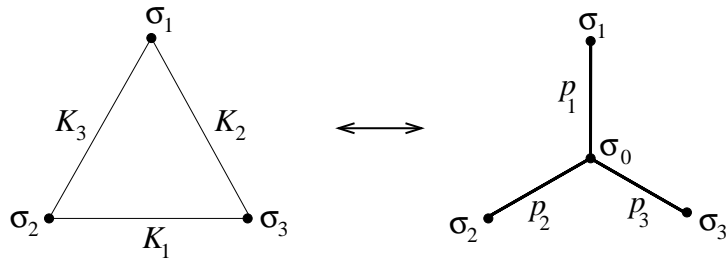}}
\end{center}
\caption{\small
The star-triangle transformation.
The coupling strengths $K_i$ of the triangle sides may be transformed into the
$p_j$ of the star legs and {\it vice versa.} 
}
\label{fig_STT}
\end{figure}

We consider an Ising model with nearest-neighbor interactions
on a triangular lattice.
Figure \ref{fig_STT} shows on the left a local configuration 
where a triangle of
spins $\sigma_1$, $\sigma_2$, and $\sigma_3$
interacts via three nearest-neighbor couplings 
$K_1, K_2,$ and $K_3$.
The expression for the partition function of this lattice may be transformed
in various ways.
It is in particular possible to ``decorate''
the lattice by inserting a central spin
$\sigma_0$ such that the triangle of couplings 
is replaced with a three-legged star of couplings $p_1$, $p_2$, and $p_3$
linking the central spin to the three others,
as shown on the right of figure \ref{fig_STT}.
The $p_j$ have well-defined expressions in terms of the $K_i$
(with $i,j=1,2,3$), and inversely.
The relevant formulas, of which a derivation may be found in Syozi
\cite{Syozi72}, will be displayed in section \ref{sec_STformula}.

The decoration transformation is fully reversible: we may 
``decimate'' the central spin in a star of pairwise interacting spins and 
thereby recover the original triangle. Applied either way, the operation is
referred to as the star-triangle (ST) transformation.

A {\it homogeneous\,} triangular or hexagonal lattice is one in which
the couplings depend only on the orientation $i$ of the bonds.
In the study of the two-dimensional (2D) Ising model
the ST transformation for the homogeneous isotropic case ($K_i=K$ and $p_i=p$
for all $i=1,2,3$)  appears in Onsager \cite{Onsager44}
and in Wannier \cite{Wannier45}. The homogeneous but anisotropic case 
(three different $K_i$ and $p_i$) seems 
to have been first exhibited by Houtappel \cite{Houtappel50}. 
This author combined the ST transformation with a high/low temperature duality
and  established the critical surfaces of the 
triangular and hexagonal Ising lattices in the $K_1K_2K_3$ and $p_1p_2p_3$ 
parameter space, respectively.
In one important line of development
the ST transformation and its generalization to the  Yang-Baxter equations
have become a cornerstone in the study of integrable models  
(see, {\it e.g.,} Perk and Au-Yang
\cite{AuYangPerk89,PerkAuYang06}).

\subsection{Star-triangle (ST) evolution}
\label{sec_startraiangleflow}

\begin{figure}[ht]
\begin{center}
\scalebox{.40}
{\includegraphics{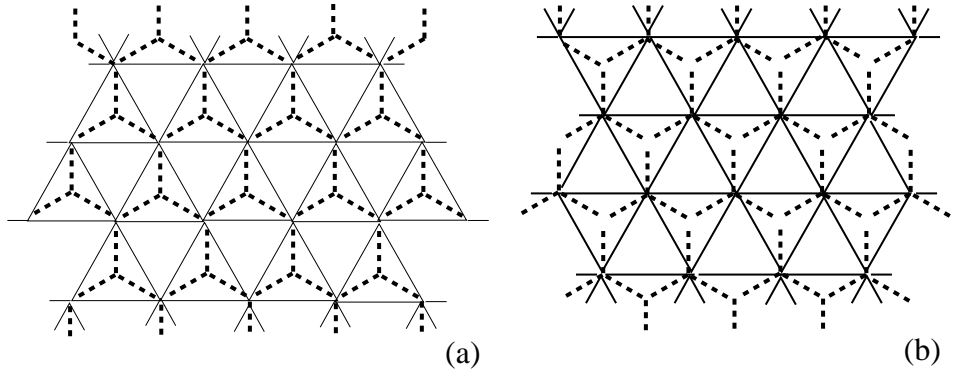}}
\end{center}
\caption{\small
(a) Thin solid lines: lattice \calTzero.
Thick dashed lines:
lattice \calHzero\ obtained from \calTzero\ by ST transformation.
(b) The same lattice \calHzero\ and, in thick solid lines,
lattice \calTone\ obtained from it by ST transformation.
}
\label{fig_STTlattice}
\end{figure}

The focus of our work is on triangular and hexagonal lattices with 
nearest-neighbor couplings $\{K_i(\vecr)\}$ that are {\it inhomogeneous,} 
{\it i.e.,} that vary with the spatial position $\vecr$.%
\footnote{The position $\vecr$ of a bond may be defined by any suitable
  convention, {\it e.g.} as the midpoint of the bond.}
Let us consider the triangular lattice \calTzero\ made up of the thin solid 
lines in figure \ref{fig_STTlattice}a and whose couplings we will denote by 
$\{K_i^{(0)}(\vecr)\}$. 
We will suppose that this lattice has no boundaries (it is infinite or
periodic in both directions).
We start by partitioning this lattice into upward pointing triangles.
Upon applying the ST transformation independently to each of these triangles
we generate a collection of upward pointing stars that together constitute
a hexagonal lattice, to be called \calHzero. It is made up of the
thick dashed lines in figure \ref{fig_STTlattice}a  
and has been reproduced in figure \ref{fig_STTlattice}b. 
We now partition this lattice in a different way, namely into 
downward-pointing stars,
and apply the ST transformation to those. This generates a collection of
downward-pointing 
triangles that together constitute a new triangular lattice that we will call
\calTone, and that is made up of the heavy solid lines in figure
\ref{fig_STTlattice}b.
The lattices \calTzero\ and \calTone\ 
differ by a translation but are geometrically identical.
If \calTzero\ is homogeneous with couplings $K_1, K_2,$ and $K_3$,
then \calTone\ will also be homogeneous and have the same couplings.
However, if \calTzero\ is inhomogeneous, then
so will \calTone\ be, but with a different set 
$\{K_i^{(1)}(\vecr)\}$ of coupling strengths.

We will view the mapping from \calTzero\ to \calTone, 
via the intermediate lattice \calHzero,  as one iteration of the ST
transformation, composed of two steps.
Successive iterations generate the sequence 
\beq
\mbox{ {\calTzero $\,\mapsto$ \calHzero $\,\mapsto$ \calTone 
$\,\mapsto$ \calHone $\,\mapsto$ \ldots $\,\mapsto$ \calTn $\,\mapsto$ \calHn
$\,\mapsto$ \ldots }, }
\label{TRIHEX}
\eeq
in which each lattice is characterized by its set of couplings,
$\{K_i^{(n)}(\vecr)\}$ or $\{p_i^{(n)}(\vecr)\}$, for $n=0,1,2,\ldots$,
and where each step, \TtoH\ or \HtoT, is followed by a repartitioning 
of the stars or of the triangles, respectively. 
We will refer to sequence (\ref{TRIHEX}) 
as the {\it ST evolution\,} of the initial lattice.%
\footnote{The ST evolution on a lattice without boundaries is fully 
reversible; our initial choice of partitioning \calTzero\ into up-triangles, 
rather than down-triangles, defines the forward direction of the evolution.
\label{foot_evol}}  
It is an {\it inhomo\-ge\-ne\-ity-driven\,} 
mapping in the space of nearest-neighbor Hamiltonians on the
triangular lattice.

\subsection{Applications}
\label{sec_applications}

ST evolution has led to various analytical and numerical applications.
It was introduced by the author {\it et al.} 
\cite{HSvL78,HSvL79}, who exploited it to construct 
an exact renormalization transformation in ``real space'' of the
2D Ising model.  
Their particular application was
based on three couplings $K_i^{(n)}(\vecr)$
that for fixed $i$ vary smoothly with the spatial position $\vecr$
and with the iteration index $n$.
ST evolution then becomes a continuous flow described by a set of 
three coupled nonlinear partial differential equations.
Subsequently the flow in the critical surface was studied in greater detail
\cite{KH79}, and similar ST evolution equations were formulated for the
$d$-dimensional Gaussian model \cite{YH79,YMH79,YMH80}, 
where stars have $d+1$ legs and triangles are 
replaced with $d$-simplices.

ST evolution has yielded new results in particular for
the class of semi-infinite 2D Ising lattices.
Under this evolution the boundary magnetization and the
boundary-spin correlation of such lattices
couple only to themselves and may be calculated numerically exactly or, 
in certain cases, analytically \cite{HvL81,Burkhardt82,Cordery82,BGHvL84,%
BurkhardtGuim84,BurkhardtGuim85,IgloiLajko96,BurkhardtGuim98,LajkoIgloi00}.
Furthermore, several of the techniques developed in this context have been 
transposed to 1D quantum chains, whether they be nonrandom
or disordered
\cite{BercheTurban90,Turbanetal94,Selkeetal97,Igloietal98,Turbanetal99,%
Karevskietal00,Turban02,Colluraetal09}.
Review articles on these semi-infinite 
models are due to Igl\'oi {\it et al.} \cite{Igloietal93} and 
Pelizzola \cite{Pelizzola97}.

\subsection{This work}
\label{sec_thiswork}

The present work is motivated by the 
absence, so far, of applications
of the ST transformation to frustrated Ising lattices, and
notably to Ising spin glasses.
We investigate here the behavior of ST evolution 
for initial triangular lattices \calTzero\ having 
a mixture of ferromagnetic (positive) 
and antiferromagnetic (negative) coupling strengths.
Such initial conditions allow for frustration, which will cause
complex couplings to be generated in the lattices \calTone, $\calT(2),\ldots$. 
We establish rules that are universally 
obeyed by the ST evolution of complex couplings.
The central result is that 
for a system without boundaries ({\it i.e.,}
which is infinite or periodic in both directions)  
the variables $S_i^{(n)}(\vecr)=1/\sinh 2K_i^{(n)}(\vecr)$ 
remain confined to the union of the real and the imaginary axis.
An analogous statement holds for the intermediate hexagonal lattices.
For a system with boundaries
the above statements, while remaining essentially valid, 
have to be formulated more carefully and require separate proof.

We are not dealing in this work with an application to any specific system.
We believe that the rules that we will describe 
are interesting in their own right 
and we expect them to play a role in future applications of the ST
transformation to specific frustrated lattices.

This paper is organized as follows.
In Section \ref{sec_STformula} we recall the basic ST transformation
formula
and state the main theorem, applicable to systems without a boundary.
In Section \ref{sec_colorevol} we define the concept of the ``color'' of a
bond. We derive rules for the evolution of the colors under iteration
of the ST transformation
and, relying on these rules, prove the main theorem.
In Section \ref{sec_boundaries} we consider lattices with boundaries and 
are led to distinguish between ``advancing'' and ``receding'' ones.
Stars and triangles located at the boundaries may be incomplete,
that is, they may miss legs or sides,
and we provide the required modifications of the ST transformation formulas
for those cases. 
At advancing boundaries the necessity of a ``protocol'' appears, 
as discussed in section \ref{sec_irrevprotocol}.
In Section \ref{sec_colorevolbound} we investigate the ST evolution of the 
bond colors in the presence of boundaries and prove the relevant 
modified theorems.
In Section \ref{sec_conclusion} we present our conclusions.

\section{The star-triangle transformation formula}
\label{sec_STformula}

When no confusion can arise we will drop the iteration index $n$
and the spatial position $\vecr$,
and we will write $\{K_1,K_2,K_3\}$ for an arbitrary triplet of couplings 
under consideration. Analogous notation will be used for the coupling 
strengths on the intermediate hexagonal lattices.

We will now connect the coupling strengths $p_j$ of three star legs
to the $K_i$ of the three corresponding triangle sides,
with $i,j=1,2,3$ indicating the bond orientations 
according to figure \ref{fig_STT}.

When dealing with a star $\{p_1,p_2,p_3\}$
we will let $S_i$ and $C_i$ stand for 
\beq
S_i = \sinh 2p_i\,, \qquad C_i = \cosh 2p_i\,, \qquad i=1,2,3;
\label{defSCstar}
\eeq
and, for reasons to become clear below,
when dealing with a triangle $\{K_1,K_2,K_3\}$
we will let these same variables stand for 
\beq
S_i = 1/\sinh 2K_i\,, \qquad C_i = \coth 2K_i\,, \qquad i=1,2,3.
\label{defSCtri}
\eeq
We may therefore denote a bond by $(S_i,C_i)$ without specifying if
it is a star leg $p_i$ or a triangle side $K_i$.
Definitions (\ref{defSCstar}) and (\ref{defSCtri}) both lead to the
relation
\beq
C_i^2 - S_i^2 =1, \qquad i=1,2,3.
\label{SCrelation}
\eeq

Let a triplet of {\it ingoing\,} bonds $(S_j,C_j)$, with $j=1,2,3$, 
be subjected to the ST transformation such that it
produces a triplet of {\it outgoing\,} bonds 
$(S_i^{\,\prime},C_i^{\,\prime})$ with $i=1,2,3$. 
After a somewhat tedious calculation one finds that 
the ST transformation takes the form \cite{Syozi72}
\beq
S_i^{\,\prime} = \frac{F^{1/2}}{S_jS_k}\,, \qquad
C_i^{\,\prime} = \frac{C_i+C_jC_k}{S_jS_k}\,, \qquad i,j,k\, \mbox{ cyclic,}
\label{STTCS}
\eeq
in which $F$ is given by
\beq
F = 2 + S_1^2 + S_2^2 + S_3^2 + 2C_1C_2C_3\, 
\label{defF}
\eeq
and one may verify that again
\beq
C_i^{\,\prime\, 2} - S_i^{\,\prime\, 2} =1, \qquad i=1,2,3.
\label{SCprimerelation}
\eeq
The advantage of equations (\ref{STTCS})-(\ref{SCprimerelation})
is that they are valid for both the step \TtoH\ and \HtoT,
which was the reason for definitions (\ref{defSCstar})-(\ref{defSCtri}).

The ST transformation leaves the partition function invariant up to a
multiplicative 
constant $G$, whose expression in terms of the $S_i$ and $C_i$ does not
concern us here. 
In equation (\ref{STTCS}) the sign of the square root $F^{1/2}$ may be chosen
according to an arbitrary convention.%
\footnote{When the outgoing triplet is a star, 
changing the sign of $F^{1/2}$ 
amounts to the redefinition $\sigma_0 \mapsto -\sigma_0$
of the decorating central spin, which leaves the partition function invariant;
when the outgoing triplet is a triangle,
it amounts to multiplying the partition
function by a phase factor.\label{footnote3}} 

\section{Bond color evolution}
\label{sec_colorevol}

We consider now the sequence of lattices (\ref{TRIHEX}) generated from a
given initial lattice \calTzero. 
The basic result of this work
refers to a lattice without boundaries, that is, which is infinite
or periodic in both directions. It may be stated as follows.\\

\noindent {\bf \sc Theorem \theexerc.} 
\setcounter{refa}{\value{exerc}} 
\stepex
{\it Let a triangular lattice \calTzero\ without boundaries
have initial couplings $K_i^{(0)}(\vecr)$
that are arbitrary positive or negative%
\footnote{We will not consider $K_i^{(0)}(\vecr)=0$, 
which would mean the absence of a coupling.}
reals.
Then under the star-triangle evolution {\rm (\ref{TRIHEX})},
represented by equations (\ref{STTCS})-(\ref{defF}),
the couplings $S_i^{(n)}(\vecr)$ and $C_i^{(n)}(\vecr)$ of the 
lattices \calTn\ and \calHn,
where $n=0,1,2,\ldots$,
remain confined to the union of the 
real and imaginary axis.}\\

In the remainder of this section we prove this theorem
via a succession of definitions, properties, and corollaries.

\subsection{Bond colors}
\label{sec_bondcolors}

We assign each bond to one of three ``color'' classes,
depending on its numerical value. 
This concept of color will be the key to proving Theorem \therefa.
\\

\begin{figure}[ht]
\begin{center}
\scalebox{.45}
{\includegraphics{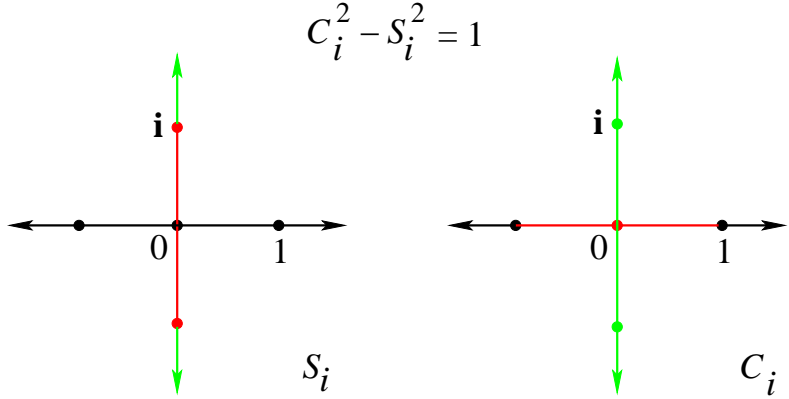}}
\end{center}
\caption{\small 
The three distinct color domains, black, red, and green,
in the complex plane of the variable $S_i$ (left) and,
under the relation $C_i^2-S_i^2=1$,
the three corresponding domains of the variable $C_i$ (right). 
}
\label{fig_regimes}
\end{figure}

\noindent {\sc Definition \theexerc.} 
\setcounter{refb}{\value{exerc}}
\stepex
{\it For each bond $(S_i,C_i)$ we
distinguish three possibilities, shown graphically in figure 
\ref{fig_regimes},

${}$\phantom{ii}{\rm (i)} ``black'' bond: $S_i$ an $C_i$ both real,
which implies $|C_i|\geq 1$;

${}$\phantom{i}{\rm (ii)} ``red'' bond: $S_i$ imaginary and $C_i$ real,
which implies $0<|S_i| \leq 1$ and $0 \leq |C_i| <1$;

${}${\rm (iii)} ``green'' bond: $S_i$ and $C_i$ both imaginary,
which implies $|S_i|>1$ and $|C_i|>0$.
}\\

\noindent
The color domains cover the union \un of the real and the imaginary axis.
When $S_i$ is not in \un, then neither is $C_i$, and {\it vice versa};
such a coupling $(S_i,C_i)$ has no color assigned to it.
For an initial lattice \calTzero\ with arbitrary positive or negative 
couplings $K_i^{(0)}(\vecr)$ all bonds are black and, 
to prove Theorem \therefa, we have to show that ST evolution
does not generate couplings outside of \un. 

\subsection{Color transformation table}
\label{sec_inoutcolors}

\begin{table}[ht]
\begin{center}
\scalebox{.25}
{\includegraphics{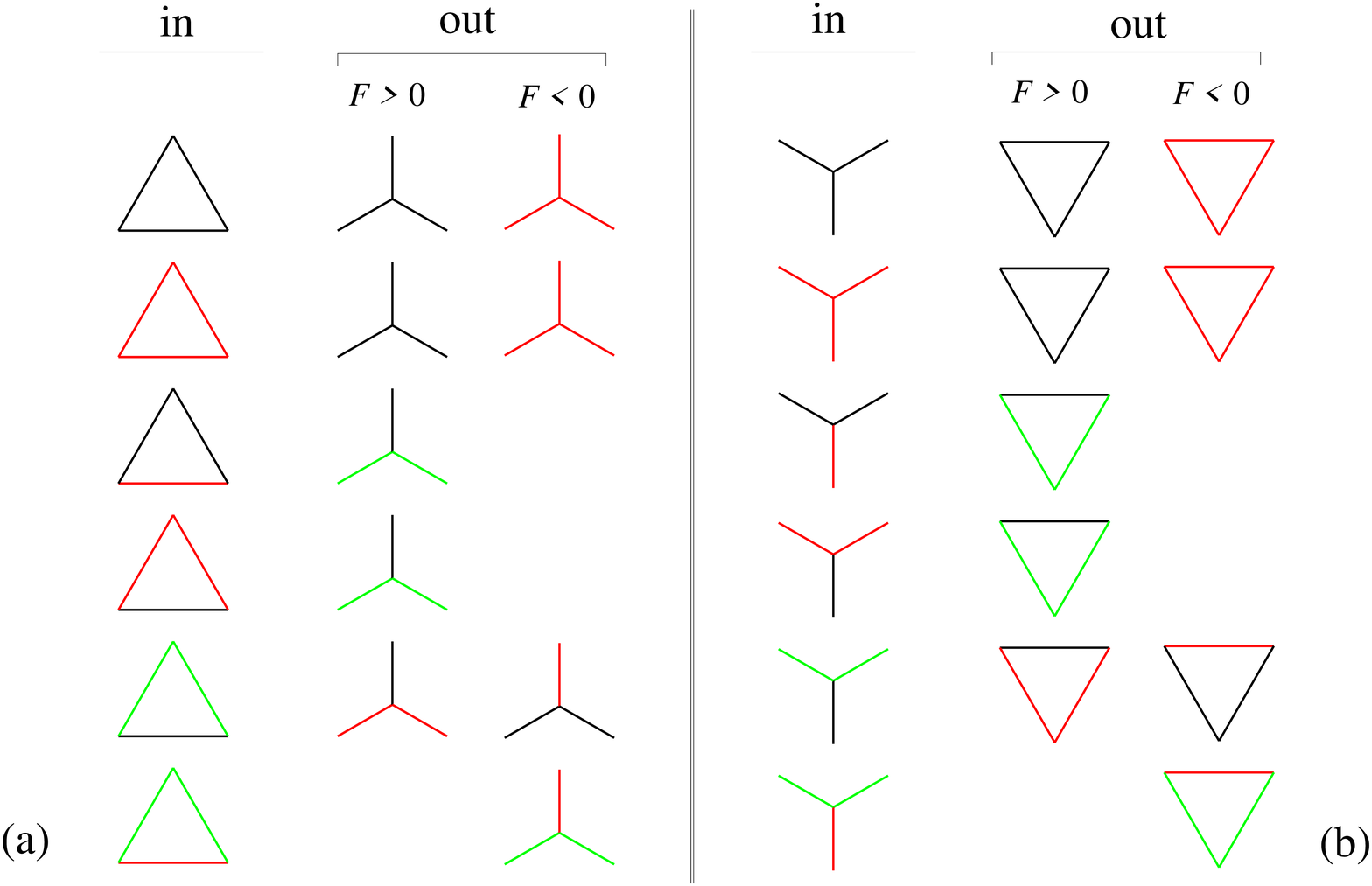}}
\end{center}
\caption{\small 
Transformation (a) from triangle to star and (b) from star to triangle. 
``In'' columns: list of possible
ingoing color configurations, up to symmetry. ``Out'' columns:
the corresponding outgoing color configurations,
which depend on the sign of $F$. 
}
\label{fig_list}
\end{table}

\begin{figure}[ht]
\begin{center}
\scalebox{.35}
{\includegraphics{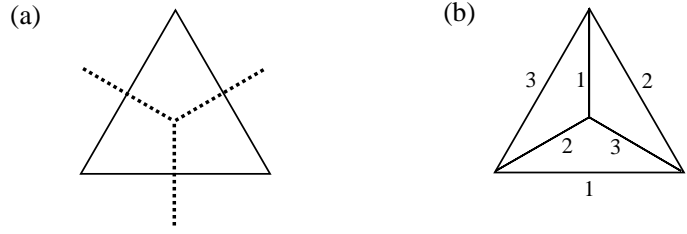}}
\end{center}
\caption{\small
(a) The three triangle sides (solid lines) with their three 
geometrically dual star legs (dashed). 
(b) For $i,j,k$ cyclic,
side $i$ of the triangle and the pair $(j,k)$ of the star legs are 
said to be adjacent; 
similarly, leg $i$ of the star and the pair $(j,k)$ of the triangle sides are
said to be adjacent. 
}
\label{fig_term}
\end{figure}

\begin{figure}[ht]
\begin{center}
\scalebox{.50}
{\includegraphics{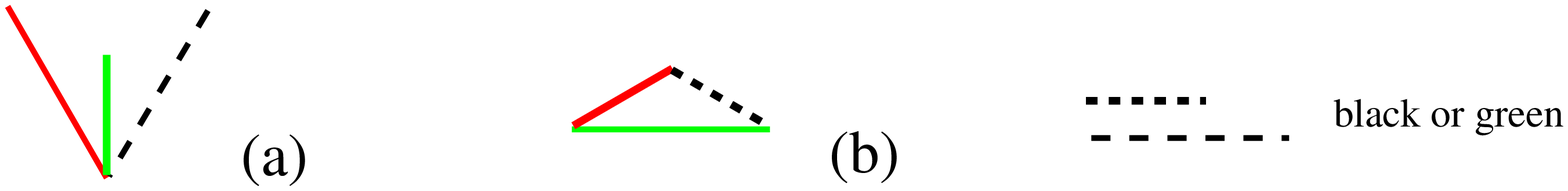}}
\end{center}
\caption{\small
(a) Illustration of Property 4. (b) Illustration of Property 5. 
}
\label{fig_prop45}
\end{figure}

We consider the ST transformation (\ref{STTCS})-(\ref{defF}) 
from three given ingoing bonds $(S_j,C_j)$,  each having one 
of the three colors black, red, or green, to three
outgoing bonds $(S_i^{\,\prime},C_i^{\,\prime})$. 
We ask, given the colors of the ingoing bonds, what 
can be said about the colors of the outgoing bonds. 
The ingoing bond triplets that concern us, up to symmetry, are
all listed in the ``in'' column of table \ref{fig_list}a
(for the steps triangle to star, \TtoH\ )
and in the ``in'' column of table \ref{fig_list}b 
(for the steps star to triangle, \HtoT\ ). 
Notably absent from the ``in'' columns are those triangles or stars 
that have an odd number (one or three) of green bonds. 
This anticipates the result, still to be
proven, that such in-triplets do not occur.

Equation (\ref{defF}) now implies
that all ingoing triplets listed in table \ref{fig_list} lead to
an $F$ that is real-valued.
Equations (\ref{STTCS})-(\ref{defF}) taken together then show that
the three outgoing bonds have well-defined colors that are fully determined
by those of the incoming bonds {\it and\,} the sign%
\footnote{The sign of $F$ cannot be determined from the ingoing
bond colors alone, but requires knowledge of the bond {\it strengths}.
These strengths therefore act as ``hidden variables'' in the bond color
evolution rules.
\label{footnote2}} 
of $F$.
It is easy to verify that this leads to the entries shown in the ``out'' 
columns in table \ref{fig_list}.

One may note that figures \ref{fig_list}a and \ref{fig_list}b are
related by duality: by this we shall mean here the operation of
replacing each bond with its geometrically dual bond
(see figure \ref{fig_term}a) while leaving the color unchanged.%
\footnote{This duality operation is not concerned with the strengths of
the bonds.} 

\subsection{Local properties}
\label{sec_local}

We still have to prove that table \ref{fig_list} actually covers all cases
that may occur.  
To that end we first observe the following two properties of ST transformation 
(\ref{STTCS})-(\ref{defF}).
\\

\noindent {\sc Property \theexerc }a. 
\setcounter{refc}{\value{exerc}}
{\it Let an ingoing triplet consist of bonds $(S_j,C_j)$, with $j=1,2,3$, 
that all have one of the colors black, red, or green.
Then $2+S_1^2+S_2^2+S_3^2$ is real and the variable $F$
moves off the real axis if and only if $C_1C_2C_3$ is pure imaginary. 
In that case the outgoing $(S_i^{\,\prime},C_i^{\,\prime})$ will not be in \un.}\\

\noindent {\sc Property \theexerc }b. \stepex
{\it An ingoing triplet of bonds all having one of the colors black, red, 
or green will generate outgoing bonds not in \un
if and only if an odd number
(that is, one or three) of the ingoing bonds is green.}\\

\noindent {\sc Proof of Properties} {\therefc}a {\sc and} {\therefc}b.  
It suffices to observe that the product $C_1C_2C_3$ is real for zero or two
green bonds, and pure imaginary for one or three green bonds.
$\square$\\  


\setcounter{refe}{\value{exerc}}
\stepex
\setcounter{refg}{\value{exerc}}
\setcounter{reff}{\value{exerc}}
\setcounter{refh}{\value{exerc}}

\noindent Three further properties will be needed. The first two of them
involve the notion of adjacency, defined in figure \ref{fig_term}b.\\

\noindent{\sc Property \therefe.} 
{\it Consider a star and a triangle associated with each other by 
  {\rm ST} transformation. A leg of the star is green if and only if exactly
  one of its two 
  adjacent triangle sides is red.}\\

\noindent{\sc Property \therefg.} 
{\it Consider a star and a triangle associated with each other by  
{\rm ST} transformation. A side of the triangle is green if and only if
exactly one of its  
two adjacent star legs is red.}\\

Properties \therefe\ and \therefg\ are illustrated in figure \ref{fig_prop45}a
and \ref{fig_prop45}b, respectively.
These two properties are related by duality.
Moreover, because of the ``if and only if,''
both properties may be applied in the step \TtoH\
and in the step \HtoT.
Their potential stems from the fact that only partial knowledge is needed
about the colors of the three ingoing bonds (star legs or triangle sides)
that participate in the transformation.
We will invoke these properties repeatedly in sections \ref{sec_theoremten},
\ref{sec_receding}, and \ref{sec_advancing}.\\

\noindent {\sc Proof of properties \therefe\ and \therefg.}
One easily verifies these properties by examining all cases in 
figures \ref{fig_list}a  and \ref{fig_list}b. 
$\square$\\

\stepex
\noindent {\sc Property \theexerc.}
\setcounter{refi}{\value{exerc}}
{\it If the number of ingoing green bonds in an {\rm ST\,} transformation
is even (that is, zero or two),
then the number of outgoing green bonds is also even.}\\

\noindent {\sc Proof of Property \therefi.}
See table \ref{fig_list}.
$\square$\\

The properties listed so far are all local:
they refer to the conversion of a local triplet of star legs
into a local triplet of triangle sides, or {\it vice versa,}
irrespective of their environment.
In the next subsection we consider the interaction between 
neighboring triplets that arises as a consequence of the 
repartitionings that follow each iteration step.

\subsection{Nonlocality}
\label{sec_nonlocality}

Each iteration of the ST transformation comprises two repartitionings of
bonds, the first one (from up-stars to down-stars) after the step
\TtoH, and the second one (from down-triangles to up-triangles) 
after the step \HtoT.
The repartitionings rearrange the bonds into new triplets,
and as a consequence the ST evolution becomes nonlocal.
In order to efficiently discuss the implications of this we introduce
two new definitions.\\

\stepex
\noindent {\sc Definition \theexerc }a.
\setcounter{refj}{\value{exerc}}
{\it A triangle in a triangular lattice
is said to be even-green \parEG\
if it contains an even number 
(zero or two) of green sides. A triangular lattice
is said to be even-green with respect to the up-triangles 
\parEGup\ if all its up-triangles are \EG;
it is said to be even-green with respect to the down-triangles 
\parEGdown\ if all its down-triangles are \EG.
The lattice
is said to be even-green \parEG\ if it is both \EGup\ and\, \EGdown.}\\

\noindent {\sc Definition \theexerc }b.
{\it A star in a hexagonal lattice
is said to be even-green \parEG\ if it contains an even number 
(zero or two) of green legs. A hexagonal lattice
is said to be even-green with respect to the up-stars 
\parEGup\ if all its up-stars are \EG;
it is said to be even-green with respect to the down-stars 
\parEGdown\ if all its down-stars are \EG.
The lattice
is said to be even-green \parEG\ if it is both \EGup\ and\, \EGdown.}\\

\noindent A lattice \calTzero\ with real couplings $K_i^{(0)}(\vecr)$
is obviously \EG. Clearly if all lattices generated from \calTzero\
in sequence (\ref{TRIHEX}) are \EG, 
then, because of properties \therefc a and \therefc b, 
$F$ cannot move off the real axis
and Theorem 1 is proved. We will therefore study 
in section \ref{sec_theoremten} the transmission of the \EG\
property under the action of the ST transformation.

We conclude this subsection with two corollaries
that state nonlocal properties of \EG\ lattices,
and with two further definitions.\\

\stepex
\noindent {\sc Corollary \theexerc }a.
\setcounter{refk}{\value{exerc}}
{\it If a hexagonal lattice without boundaries
is \EG, then its green bonds constitute a
collection of loops, and inversely. These loops are necessarily
non-self-intersecting and not mutually intersecting}.\\

\noindent {\sc Proof of Corollary \theexerc }a.
Let there be an up-star two of whose bonds, say $\alpha$ and $\beta$,
are green.
Consider one of these bonds, say $\beta$. It is also part of a down-star,
which because of the lattice being \EG\
must contain a unique other green bond, say $\gamma$.
But $\gamma$ is again also part of an up-star, {\it etc.} 
The sequence of bonds $\alpha$, $\beta$, $\gamma$, \ldots, terminates only
when it bites its own tail and then forms a loop.
The inverse is immediate: if there were a star with one or three 
green legs, then these legs would not form loops.
$\square$\\

\noindent {\sc Corollary \theexerc }b.
{\it If a triangular lattice without boundaries
is \EG, then the duals of its green bonds 
constitute a collection of loops on its dual hexagonal\,}%
\footnote{This hexagonal lattice serves merely to 
generate a picture; it 
is distinct from the hexagonal lattices
\calHn\ that appear intermediately in the iteration steps.\label{footnote1}}
{\it lattice, and inversely. 
These loops are necessarily non-self-intersecting and not
mutually intersecting.}\\

\noindent {\sc Proof of Corollary \theexerc }b.
Each bond of the dual lattice has the color of the corresponding primal bond.
This dual lattice thus colored is hexagonal and,
since under this duality \EG\ triangles yield \EG\ stars, it is \EG.
This reduces the proof to the case of Corollary \therefk a.
$\square$\\

\stepex 
\noindent {\sc Definition \theexerc }a.
\setcounter{refp}{\value{exerc}}
{\it A site of a triangular lattice is said to be even-red \parER\ 
if it has an even number of red bonds (zero, two, four, or six)
connected to it. A triangular lattice is said to be \ER\ 
if all its sites are \ER.}\\

\noindent {\sc Definition \theexerc }b.
\setcounter{refq}{\value{exerc}}
{\it An elementary six-legged loop on a hexagonal lattice is said to be 
even-red \parER\ if it contains an even number (zero, two, four, or six)
of red legs. A hexagonal lattice is said to be \ER\ if
all its elementary loops are \ER.}\\


We are now prepared for the proof of the main theorem.\\

\stepex
\setcounter{refl}{\value{exerc}}

\subsection{Main theorem}
\label{sec_theoremten}

\begin{figure}[ht]
\begin{center}
\scalebox{.40}
{\includegraphics{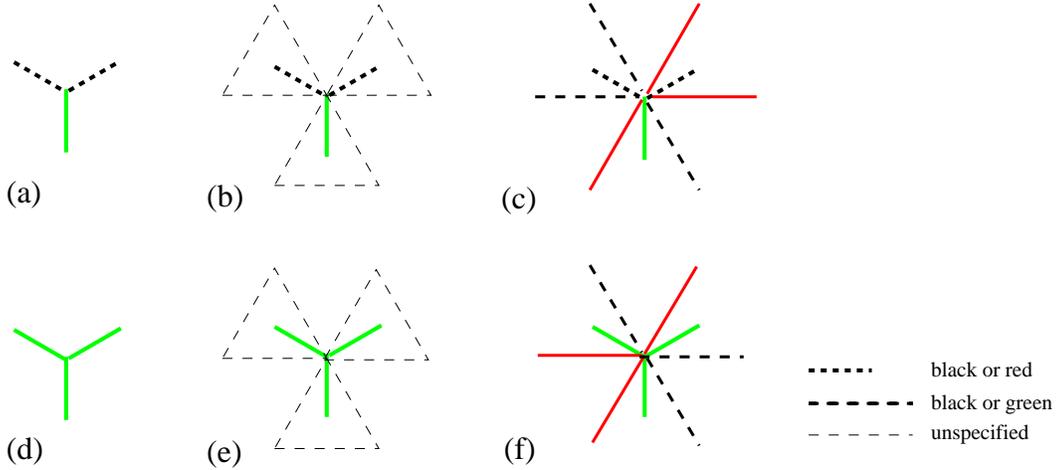}}
\end{center}
\caption{\small
Illustration of Step 1 in the proof of Theorem \therefl.
(a) An offending down-star on lattice \calHn, 
having an odd number of green legs.
(b) Thin dashed lines 
indicate the three up-triangles on lattice \calTn\
that generated the three star legs.
(c) One subset of several possible color configurations of the triangle
sides that would lead to the offending down-star; these color
configurations all violate the \ER\ property of \calTn, 
thus showing that the down-star in (a) cannot
occur. Note that the only triangle sides whose colors matter are the six
that join the star center. 
Figures (d), (e), and (f) are analogous to (a), (b), and (c), but 
for a different offending down-star. 
}
\label{fig_proofTH}
\end{figure}

\begin{figure}[ht]
\begin{center}
\scalebox{.40}
{\includegraphics{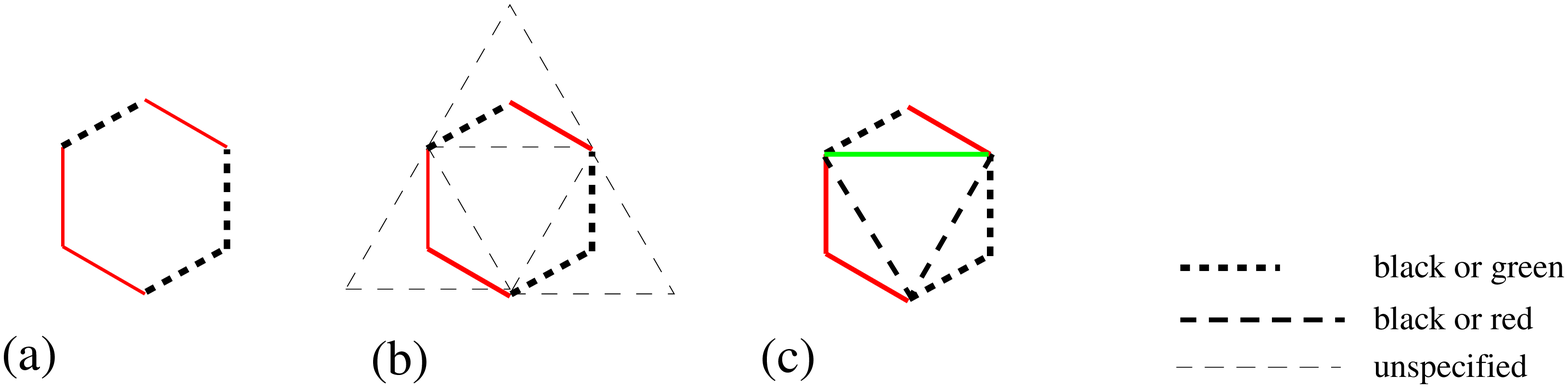}}
\end{center}
\caption{\small
Illustration of Step 1 in the proof of Theorem \therefl.
(a) An offending hexagonal loop on lattice \calHn, 
having an odd number of red legs.
(b) Thin dashed lines indicate the three up-triangles on lattice \calTn\
that generated the six legs of the loop.
(c) One subset of several possible
color configurations of the three sides of the central down-triangle 
that would lead to the offending hexagonal loop; these configurations 
all violate the \EGdown\ property, thus showing that the 
hexagonal loop in (a) cannot occur.
Note that there are only three triangle sides whose colors matter.
}
\label{fig_proofTHbis}
\end{figure}

\noindent {\sc Theorem \theexerc.} \stepex
{\it Let an initial lattice \calTzero\ be \EG\ and \ER.
Then under the {\rm ST\,} evolution {\rm (\ref{TRIHEX})}
all subsequent lattices {\rm\calHn } for $n=0,1,2,\ldots$ 
and\, {\rm\calTn}\, for $n=1,2,\ldots$ are \EG\ and \ER.}\\

\noindent {\sc Proof of Theorem \therefl.}
The proof proceeds by induction and amounts to checking all 
in- and outgoing color combinations that may occur.
Step 1 transfers the 
properties of being \EG\ and \ER\ from \calTn\
to \calHn, and Step 2 transfers them from \calHn\ to \calTnpone.\\

{\it Step 1:} \TtoH. --- 
Let \calTn\ be both \EG\ and \ER\ and suppose that \calHn\ is not. 
We will show that this leads to a contradiction.

Suppose first that \calHn\ is not \EG.
Since the \EGup\ property of \calHn\ follows from the \EGup\
property of \calTn\, we must then have that \calHn\ is not \EGdown.
This means there is an ``offending'' down-star 
of the type shown in figure \ref{fig_proofTH}a 
(possibly up to a rotation) or in \ref{fig_proofTH}d, 
having either one or three green bonds.
The dashed black bonds in figure \ref{fig_proofTH}a
stand for bonds that may be either red or black.
We now trace these two configurations one step back. 
The three legs were generated by three different up-triangles
of the lattice \calTn, shown by thin dashed lines
in figures \ref{fig_proofTH}b and \ref{fig_proofTH}e.
Each of the three star legs has two adjacent triangle sides, 
and these six sides come together at the center of the down-star.
Now by Property \therefe, a leg can be green if and only if among
the two adjacent sides of its generating triangle
there is an odd number (necessarily one) of red sides. 
The number of red triangle sides joining in the star center is therefore:
in figure \ref{fig_proofTH}c, the sum of an odd integer (one) and two even
integers (equal to zero or to two)
and in figure \ref{fig_proofTH}f, the sum of three odd integers 
(each equal to one).
Hence this number is odd, which contradicts 
the premise that \calTn\ is \ER. It follows that \calHn\ is \EG.

Suppose, secondly, that \calHn\ is not \ER. 
This means there is an offending elementary six-legged loop that has an odd
number of red legs. An example of such a loop is shown in
figure \ref{fig_proofTHbis}a.
These legs were generated pairwise in three independent up-triangles of
\calTn, shown by thin dashed lines in figure \ref{fig_proofTHbis}b.
They together enclose a central down-triangle each of whose sides
is adjacent to two consecutive legs of the loop.
Now by Property \therefg,
such a leg pair has an odd number (hence exactly {\it one}) 
of red legs if and only if the adjacent side of the generating triangle
is green.
Therefore, an odd number of red legs in the loop 
implies that the central down-triangle has an odd number of
green sides. This contradicts the premise
that \calTn\ is \EG. It follows that \calHn\ is \ER.

Hence \calHn\ is both \EG\ and \ER.\\

\begin{figure}[ht]
\begin{center}
\scalebox{.40}
{\includegraphics{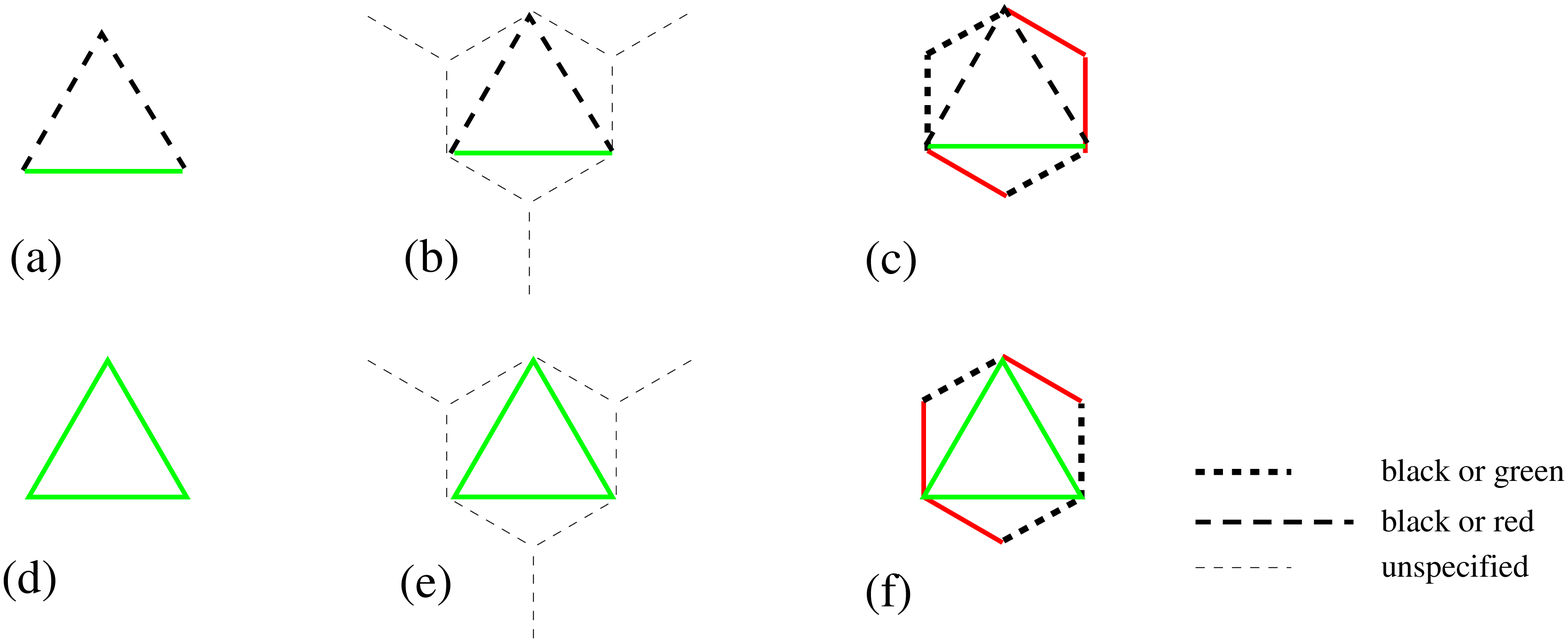}}
\end{center}
\caption{\small
Illustration of Step 2 in the proof of Theorem \therefl.
(a) An offending up-triangle on lattice \calTnpone, 
having an odd number of green legs.
(b) Thin dashed lines indicate the three down-stars on lattice \calHn\
that generated the three triangle sides.
(c) One subset of several possible color configurations of the star
legs that would lead to the offending up-triangle; 
these configurations all violate the \ER\ property of \calHn, 
thus showing that the up-triangle in (a) cannot
occur. Note that the only six star legs whose colors matter are those
that are part of the hexagonal loop around the triangle in (a). 
Figures (d), (e), and (f) are analogous to (a), (b), and (c), but 
for a different offending up-triangle.
}
\label{fig_proofHT}
\end{figure}

\begin{figure}[ht]
\begin{center}
\scalebox{.40}
{\includegraphics{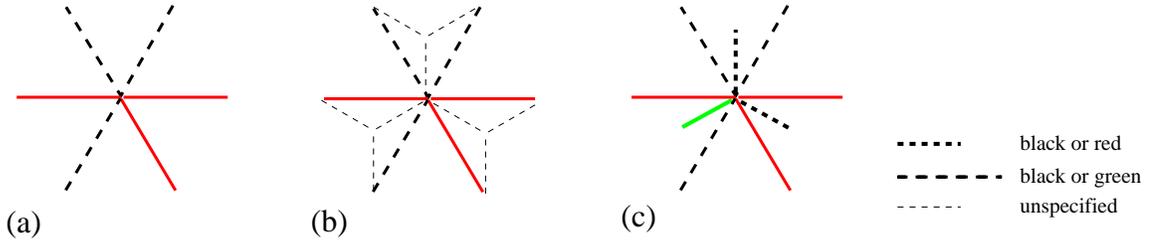}}
\end{center}
\caption{\small
Illustration of Step 2 in the proof of Theorem \therefl.
(a) An offending site on \calTnpone, joined by an odd number of red triangle
sides. 
(b) Thin dashed lines indicate the three down-stars on lattice \calHn\
that generated the six triangle sides joining the offending site.
(c) One subset of several possible
color configurations of the three legs of the central down-star
that would lead to the offending site; these configurations 
all violate the \EGup\ property of \calHn, thus showing that the 
site in (a), where three triangle sides join, cannot occur.
Note that the only three star legs whose colors play a role
are those shown in (c).
}
\label{fig_proofHTbis}
\end{figure}

{\it Step 2:} \HtoT. ---
Let \calHn\ be both \EG\ and \ER\ and suppose that \calTnpone\ is not. 
We will show that this leads to a contradiction.

This step is related to Step 1 by symmetry under duality.
We nevertheless repeat the reasoning because it will be of help
in the next section in discussing
a system with boundaries, for which the same duality is absent. 
Local color configurations relevant to this step
are shown in figure \ref{fig_proofHT}.

Suppose first 
that \calTnpone\ is not \EG. Since its \EGdown\ property follows from
the \EGdown\ property of \calHn, we must then have that \calTnpone\ is not
\EGup. This means there must be an offending up-triangle which, 
possibly up to a rotation, is of the type shown in figure 
\ref{fig_proofHT}a or \ref{fig_proofHT}d, having either
one or three green sides. Again, the dashed black lines in 
\ref{fig_proofHT}a stand for bonds that
may be either red or black. The three triangle sides were generated by three
different down-stars of the lattice \calHn, shown in
figures \ref{fig_proofHT}b and \ref{fig_proofHT}e 
as thin dashed lines.
Each of the three triangle sides has two adjacent legs, and these six
legs constitute an elementary loop  of \calHn. 
Now when Property \therefh\ 
is applied to the three down-stars, 
we find that this elementary loop
contains an odd number of red legs
as shown in \ref{fig_proofHT}c and  \ref{fig_proofHT}f.
This contradicts the premise that \calHn\
is \ER. Hence \calTnpone\ is \ER.

Suppose, secondly, that \calTnpone\ is not \ER.
Then it must have an offending site 
where an odd number of red triangle sides come together.
An example of such a is shown in figure \ref{fig_proofHTbis}a.
The six sides were generated pairwise
by three distinct down-stars of \calHn,
shown by thin dashed lines in figure \ref{fig_proofHTbis}b. 
Each down-star contributes two sides, 
of which, by Property \therefe, exactly
one is red if and only if the adjacent star leg is green.
Since there is an odd number of red sides, there must also be an odd number of
green star legs joining at the same site.
This contradicts the premise that \calHn\ is \EG.
Hence \calTnpone\ is ER.\\

Upon combining the results of Steps 1 and 2
we conclude that \calTnpone\ is both \EG\ and \ER, which proves Theorem 
\therefl.
$\square$\\

\noindent We now return to Theorem 1.\\

\noindent {\sc Proof of Theorem 1.}
A triangular lattice that has only positive or negative couplings
has only black bonds and is therefore
trivially EG and ER. Hence Theorem 1 is the special case of Theorem \therefl\
that applies to physical systems.
$\square$

\subsection{Numerical evaluation}
\label{sec_numerical}

\begin{figure}[ht]
\begin{center}
\scalebox{.50}
{\includegraphics{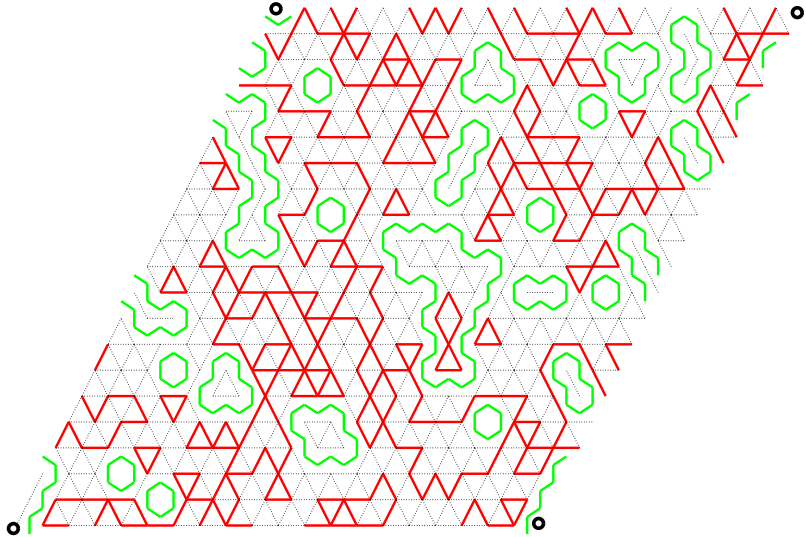}}
\end{center}
\caption{\small
Snapshot of the bond coloring of lattice $\calT(100)$.
The lattice is periodic in two directions, the four encircled 
sites being identical.
The figure was obtained 
from an initial lattice \calTzero\ with random coupling strengths
$S_i^{(0)}(\vecr)=1/\sinh 2K_i^{(0)}(\vecr)$ drawn independently for each bond 
from a uniform distribution on $[-0.5,0.5]$.
For greater clarity,
instead of showing the green bonds of $\calT(100)$ themselves, 
the figure shows their duals. See text.
}
\label{fig_snap}
\end{figure}

We performed the ST evolution numerically for a triangular lattice 
of $20 \times 20$ sites with periodic boundary conditions. 
The initial lattice $\calT(0)$ has coupling strengths $K_i^{(0)}(\vecr)$
drawn randomly and independently from a distribution symmetric about zero,
{\it i.e.,} the system is a standard 2D symmetric Ising spin glass.
Although initially all bonds are black, the frustration
of a fraction of the elementary triangles
leads to the appearance of red and green bonds after the first iteration.
Figure \ref{fig_snap} shows the bond colors of  
lattice $\calT(100)$, obtained after $100$ iterations.
For greater clarity it does not show the green triangle sides themselves, 
but rather their duals as defined by 
figure \ref{fig_term}a, which are star legs.
The green star legs are seen to form loops in agreement with
Corollary \therefk b\ and with the lattice being \EG.
Moreover, each lattice site is joined by an even number, possibly zero, 
of red bonds, in agreement with Definition \therefp a\
and with the lattice being \ER. 

We have briefly considered the numerical errors incurred under ST
transformation. 
We applied to an initial lattice $\calT(0)$ of size
$20 \times 20$ first $n$ forward iterations so as to obtain
$\calT(n)$, and then $n$ backward iterations
leading to a final lattice $\calT^{\,\prime}(0)$; 
in the absence of numerical errors 
we would have $\calT^{\,\prime}(0) = \calT(0)$.
We took as our criterion an average discrepancy of one percent
between corresponding bond strengths of the initial and the final lattice.
For a symmetric distribution (whether dichotomic, block, or Gaussian)
of the initial coupling strengths $S_i=S_i^{(0)}(\vecr)$ we found that 
under this criterion, and using standard double precision,
we could safely iterate up to but not beyond 
$n\approx 200$ for $\langle S_i\rangle^{1/2}=\pm 2.0$ (high temperature)
and up to but not beyond
$n\approx 100$ for $\langle S_i\rangle^{1/2}=\pm 0.2$ (low temperature).
This accuracy for the symmetric spin glass appears to be low compared to what
one obtains 
in the ferromagnetic regime (all couplings random but positive);
in that regime we found that, depending on the exact parameters, $n$ may
easily take values as high as $10^4$ or $10^5$.

There is a physical explanation for the strongly deteriorated
accuracy in the case of a symmetric spin glass, 
namely the cancellations that occur 
between energy and entropy in the calculation of the system free energy.
In the spin glass phase, {\it i.e.} below the critical spin glass 
temperature $T_g$, such cancellations were explained by Fisher and Huse  
\cite{FisherHuse86,FisherHuse88} in terms of their ``droplet model.''
Thill and Hilhorst \cite{ThillHilhorst96} showed the relevance of
similar cancellations also in the temperature region closely {\it above\,}
the critical temperature, which
for the 2D spin glass has the value $T_g=0$.
In the limit $T\to T_g$ these cancellations lead to the 2D spin glass 
having two nontrivial exponents, 
the temperature exponent $y$ and the chaos exponent $\zeta$.
One may speculate that
there is a way, still to be devised, to calculate these exponents
with the aid of the ST transformation.

\section{ST evolution in the presence of boundaries}
\label{sec_boundaries}

The preceding sections apply to a lattice that does not have boundaries,
as happens when it is infinite or when suitable boundary conditions 
({\it  e.g.} periodic ones) are imposed.
However, some of the most interesting applications of the ST transformation
have involved lattices with boundaries, 
and such systems must be considered separately.
When a triangular (a hexagonal) lattice with a boundary
is partitioned into up-triangles or
down-triangles (into up-stars or down-stars), there remain at the boundary
triangles (stars) that are incomplete, {\it i.e.}, that do not have the full
set of three bonds. 
We will show how the ST transformation applies to such incomplete triangles
and stars, and will conclude that
modified versions of Theorem 1 remain valid.

\begin{figure}[ht]
\begin{center}
\scalebox{.40}
{\includegraphics{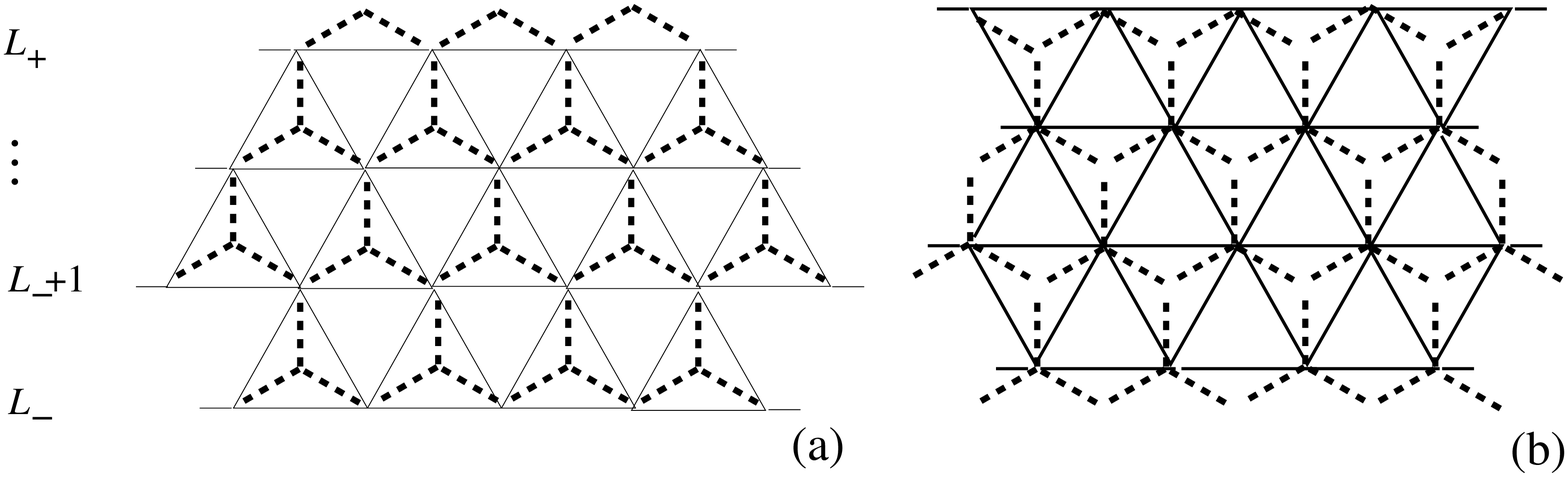}}
\end{center}
\caption{\small
(a) Thin solid lines: lattice \calTn, filling a horizontal slab 
of vertical width $L=L_+-L_-$.
Horizontally it may be infinite or periodic.
Thick dashed lines:
lattice \calHn\ obtained from \calTn\ by ST transformation.
(b) The same lattice \calHn\ and, in thick solid lines, lattice \calTnpone\
obtained from it by ST transformation.
}
\label{fig_horbound}
\end{figure}

\subsection{Advancing and receding boundaries}
\label{sec_advrec}

The simplest example of a lattice with boundaries 
is the triangular lattice \calTn\ shown  
in figure \ref{fig_horbound}a. 
It fills a slab of finite width $L=L_+-L_-$ in the vertical direction 
and may be either infinite 
or periodic in the horizontal direction.
In Step 1 of the $(n+1)$th ST iteration  
it gets transformed into the hexagonal lattice \calHn\
shown in thick dashed lines in figure \ref{fig_horbound}a
and reproduced in \ref{fig_horbound}b.
In Step 2 of the same iteration it gets transformed into the 
triangular lattice \calTnpone\ indicated by thick solid lines
in figure \ref{fig_horbound}b.
Lattice \calTnpone\ is a slab 
that is geometrically identical to \calTn\
but its bond strengths have changed
and it has moved in the vertical direction by a distance
equal to one third of the vertical separation between two successive rows.
The ST transformation in such slabs (but restricted to ferromagnetic 
interaction) was studied by Burkhardt and Guim 
\cite{BurkhardtGuim85,BurkhardtGuim98}
and by Lajk\'o and Igl\'oi \cite{LajkoIgloi00}.
This transformation involves two special types of ingoing bond
configurations:

${}$ \phantom{i}(i) In Step 1 the bonds in the top row (row $L_+$) 
of the lattice are isolated triangle sides.
Decorating such a single-sided ``triangle'' amounts to
inserting a new spin variable between the two spins joined by that side, 
which replaces the single triangle side by a pair of star legs.

${}$ (ii) In Step 2
each spin in the bottom row (row $L_-$) is connected to only two 
star legs; it is the center of an incomplete down-star.
Decimating this spin produces a single triangle side.

Clearly the two borders are not equivalent:
the upper border advances (moves into previously empty space),
and the lower border recedes (withdraws from previously occupied space).%
\footnote{The inequivalence obviously results from our choice of the forward
direction of the ST transformation; cee footnote \ref{foot_evol}.}
Obviously by setting $L_+=\infty$ (or $L_-=\infty$) we obtain a 
half-infinite system with only a receding (or only an advancing) boundary.

Below we will 
first establish the special ST transformation formulas applicable
at the boundaries and then study the corresponding 
evolution of the bond colors.

\subsection {Boundary star-triangle transformation}
\label{sec_boundarySTT}

At the boundaries the symmetry
argument that motivated definitions (\ref{defSCstar})-(\ref{defSCtri})
is of very limited use.
We will therefore in this section, in deviation from those definitions, 
reserve the notation $S_i$ and $C_i$ for
the bonds of a two-legged star, whence
\beq
S_i = \sinh 2p_i\,, \qquad C_i = \cosh 2p_i\, , \qquad i=2,3,
\label{defSC}
\eeq
whereas for an isolated triangle side we adopt the notation 
\beq
X_1 = 1/\sinh 2K_1\,, \qquad Y_1 = \coth 2K_1\,.
\label{defXY}
\eeq
Obviously
\beq
C_i^2-S_i^2 = 1, \qquad Y_1^2-X_1^2 = 1.
\label{XYrelation}
\eeq
We will consider separately
the incomplete decimation (star-to-triangle, \HtoT) and the 
incomplete decoration (triangle-to-star, \TtoH).

\begin{table}[ht]
\begin{center}
\scalebox{.35}
{\includegraphics{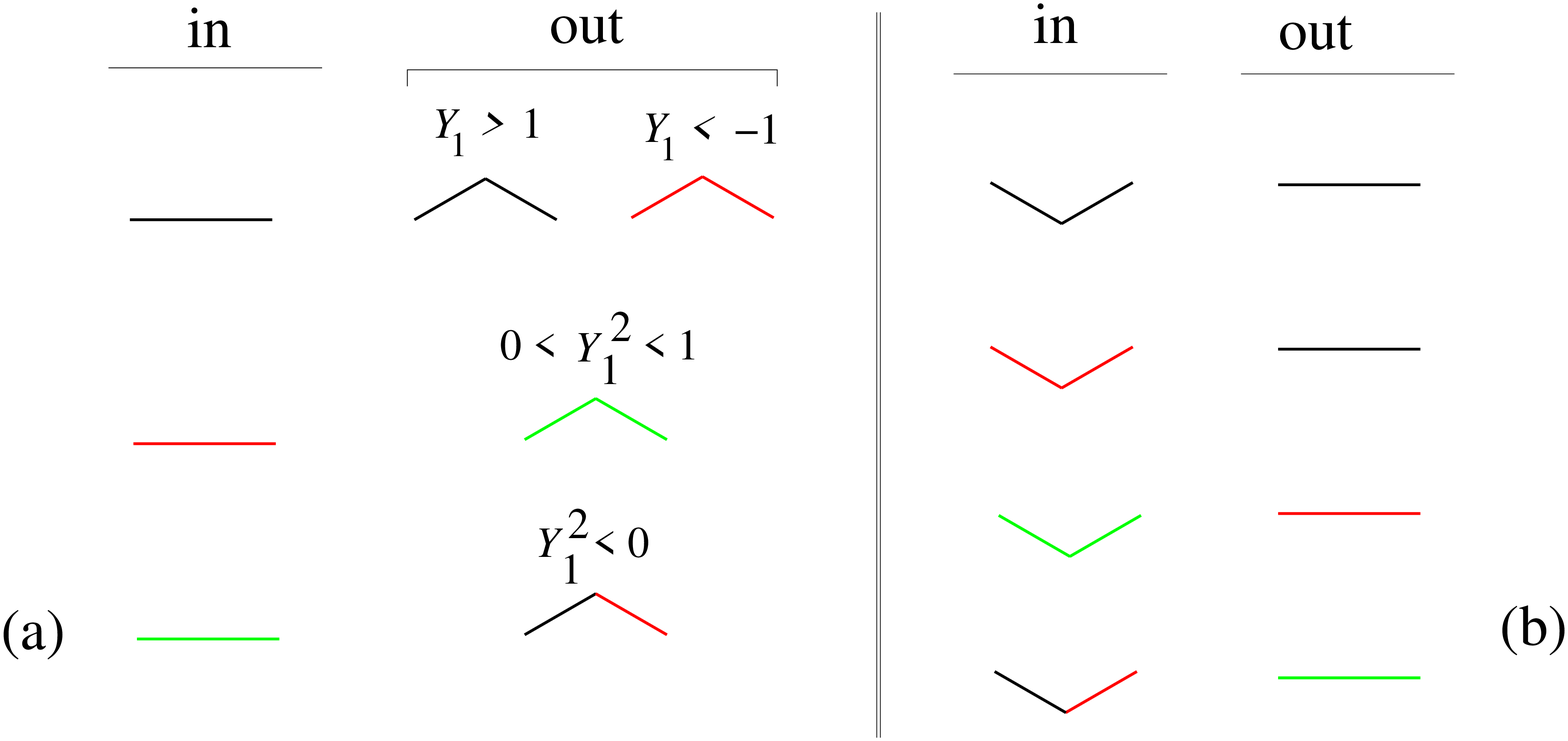}}
\end{center}
\caption{\small
Special cases at boundaries. Transformation (a) from a single triangle side 
at an advancing boundary into an incomplete two-legged star 
and (b) from an incomplete two-legged star at a receding boundary into 
a single triangle side.
}
\label{fig_listspec}
\end{table}



\subsubsection{ From a two-legged star to a single triangle side}
\label{sec_STTevolrec}

At a receding boundary we have to deal with incomplete two-legged stars.
Decimating the central spin of such a star
generates a single triangle side. Its strength $K_1$
follows from the general case 
[equations (\ref{defSCstar}) and (\ref{STTCS})] 
when we set $p_1=0$ and hence $S_1=0$ and $C_1=1$.
We find, now using  the notation of equation (\ref{defXY}), that
\beq
X_1 = \frac{C_2 + C_3} {S_2S_3}\,, \qquad 
Y_1 = \frac{1 + C_2C_3}{S_2S_3}\,. 
\label{SCtoXY}
\eeq
With the aid of (\ref{SCtoXY}) and Definition \therefb\
one easily finds the color transformation rules 
listed in table \ref{fig_listspec}b.
The color of the triangle side generated follows
deterministically by the colors of the two ingoing star legs.
Absent from the ``in'' column are leg pairs of which exactly 
one leg is green.
Such an ingoing pair would lead to an outgoing bond not in \un, and
below we will show that such a pair cannot occur.

\subsubsection{From a single triangle side to a two-legged star}
\label{sec_STTevoladv}

At an advancing boundary
we have to deal with single triangle sides.
Decorating such a side by a central spin
generates a two-legged star.
This transformation may at first sight appear as the mere inverse of equation 
(\ref{SCtoXY}); however, since (\ref{SCtoXY}) maps 
the two bonds $(S_2,C_2)$ and $(S_3,C_3)$ onto the single bond $(X_1,Y_1)$,
the inverse is not unique. This point requires a more detailed discussion,
which becomes easier
when with the aid of some simple hyperbolic trigonometry
we write (\ref{SCtoXY}) in the alternative form%
\footnote{A useful relation is $\tanh x = -1/\sinh 2x + \coth 2x$.}
\beq
\tanh K_1 = \tanh p_2 \,\tanh p_3\,,
\label{ptoK}
\eeq 
a relation well-known in 1D decimation.
For given $K_1$ this equation admits for the pair $(p_2,p_3)$
a two-dimensional continuum of solutions that may be parametrized as
\beq
\tanh p_2 = z(\tanh K_1)^{1/2}, \qquad \tanh p_3 = z^{-1}(\tanh K_1)^{1/2}, 
\label{ptoKsol}
\eeq
where $z$ is any nonzero complex number and we may arbitrarily choose
either of the square roots of $\tanh K_1$.
For general $z$ this would put $S_2, C_2, S_3$, and $C_3$ outside of \un.
We will not explore here the consequences of such a general choice of $z$,
but simply {\it impose\,} that the $S_i$ and $C_i$ remain confined to \un.
This still leaves, for given $K_1$, 
a one-dimensional continuum of possible solutions $(p_2,p_3)$ to (\ref{ptoK}),
out of which we will select a convenient one.
We have to distinguish two cases, depending on the color of the ingoing 
bond $K_1$.\\

{\it $K_1$ is black or red.} --- 
Let the triangle side $K_1$ be given. 
We select a unique solution by imposing the additional restriction 
$p_2=p_3$. This means setting $z=1$ in equation
(\ref{ptoKsol}) and leads to
\beq
\tanh p_2 = \tanh p_3  = (\tanh K_1)^{1/2}.
\label{symmsol}
\eeq 
We may now revert to the variables of equations (\ref{defSC})-(\ref{defXY}). 
With the aid of the relation $\tanh K_1 = Y_1-X_1$ and some further algebra
we find that (\ref{symmsol}) leads to
\beq
S_2 = S_3 = \frac{\sqrt{2(Y_1-X_1)}}{1-Y_1+X_1}\,, \qquad
C_2 = C_3 = \frac{Y_1+1}{X_1}\,,
\label{symmsolstan}
\eeq
which constitutes a solution to (\ref{ptoK}), and in which the sign of the
square root may be chosen arbitrarily.%
\footnote{Changing this sign merely redefines the sign of the decorating
spin, as in footnote \ref{footnote3}.} 
With the aid of (\ref{symmsolstan}) 
one easily finds the transformation rules for the colors
listed in table \ref{fig_listspec}a.
The two cases with black or red $K_1$ 
correspond to $Y_1$ being real with $Y_1^2>1$ or $0<Y_1^2<1$, respectively.
Moreover, for black $K_1$ the two subcases 
$Y_1>0$ and $Y_1<0$ have to be distinguished.%
\footnote{The sign of $Y_1$ cannot be determined from the ingoing bond colors
alone, but requires knowledge of the bond {\it strengths}. 
{\it Cf.} footnote \ref{footnote2}.}\\

{\it $K_1$ is green.} ---
When the ingoing triangle side $K_1$ is green,
then $Y_1$ and $X_1$ are pure imaginary.
If we were again to set $z=1$, whence as before solution (\ref{symmsolstan}), 
the couplings $S_i$ and $C_i$ $(i=2,3)$
would be outside \un, against our wish. 
We therefore look for a different solution.

We note that in this case $K_1$ is pure imaginary, too, 
and we therefore set $K_1 = \ii L_1$, the relevant values being 
$-\tfrac{\pi}{4}<L_1<\tfrac{\pi}{4}$.
Preparing now for the symmetry between $p_2$ and $p_3$
to be broken we let $p_3 = \ii q_3$ 
so that equation (\ref{ptoK}) becomes
\beq
\tan L_1 = \tanh p_2 \tan q_3\,. 
\label{asymmeq}
\eeq
We select the solution
\beq
\tanh p_2 = \tan q_3 = (\tan L_1)^{1/2},
\label{asymmsol}
\eeq
which amounts to setting $z=\ee^{{-\rm i}\pi/4}$ in (\ref{ptoKsol})
and is such that $\tanh p_2$ and $\tan q_3$ are either both real or both
imaginary. 
When reverting to the variables of equations (\ref{defSC})-(\ref{defXY})
and with the aid of the relation $\tanh K = -\ii(Y_1-X_1)$
we find that (\ref{asymmsol}) leads to
\begin{subequations}\label{SCtwothree}
\beq\label{SCtwo}
S_2 = \frac{ 2\sqrt{-\ii(Y_1-X_1)} }{1+\ii(Y_1-X_1)}\,, \qquad
C_2 = \frac{ 1-\ii(Y_1-X_1)       }{1+\ii(Y_1-X_1)}\,, 
\eeq
\beq\label{SCthree}
S_3 = \frac{ 2\ii\sqrt{-\ii(Y_1-X_1)} }{1-\ii(Y_1-X_1)}\,, \qquad
C_3 = \frac{ 1+\ii(Y_1-X_1)          }{1-\ii(Y_1-X_1)}\,, 
\eeq
\end{subequations}
which constitutes a solution to (\ref{ptoK}).
Since $\ii(Y_1-X_1)$ is real, so are $C_2$ and $C_3$, which moreover are seen
to satisfy the relation $C_2C_3=1$.
For $\ii(Y_1-X_1)<0$ we have that $S_2$ is real and $S_3$ imaginary,
and for $\ii(Y_1-X_1)>0$ the opposite is the case.
This means that among $p_2$ and $p_3$ one is black and the other one red.
Hence equations (\ref{SCtwo}) and (\ref{SCthree}) 
allow us to establish the last line in table \ref{fig_listspec}a.

We end this subsection with a remark about the symmetry
between $p_2$ and $p_3$. 
Setting $z=\ee^{{-\rm i}\pi/4}$ in (\ref{ptoKsol})
leads to solution (\ref{asymmsol}), which breaks this symmetry. 
Setting alternatively $z=\ee^{+\ii\pi/4}$ breaks the symmetry the other way 
around and exchanges in equations (\ref{SCtwothree})
the expressions for $(S_2,C_2)$ with those for $(S_3,C_3)$,
that is, it interchanges the red and the black star leg. 
There is no reason why one choice would be preferable over the other.
In the next subsection we address the consequences of this remaining freedom. 

\subsection{Irreversibility and protocols}
\label{sec_irrevprotocol}

The choice between the two symmetry-related solutions of the preceding
subsection may be made independently for each
triangle side in the advancing boundary, 
and independently at Step 1 of each iteration. 
Clearly in order to fully define ST evolution of a lattice with an 
advancing boundary, we have to specify a protocol.
We may, for example, assign probabilities $\eta$ and $1-\eta$,
with $0\leq\eta\leq 1$, to the two possibilities, and will refer
to this procedure as the ``random protocol.''
In the numerical example of section \ref{sec_numericalbound} below
we have chosen the ``uniform random protocol'' $\eta=1/2$. 
Protocols based on values other than $z=\ee^{\pm\ii\pi/4}$ 
are easily imagined and some may be better than
others for specific applications; 
we have not pursued that line of investigation. 

When a lattice \calTzero\ with an advancing boundary is submitted to $n$
iterations of the ST evolution under any protocol and yields \calTn,
then if we invert the evolution
the advancing boundary becomes a receding one,
the effect of the protocol is wiped out again,
and the initial state \calTzero\ is reached.

At a receding boundary the ST evolution is fully deterministic.
However, it is irreversible: 
its inverse, under which the boundary becomes advancing,
is not uniquely defined and the initial state \calTzero\ cannot
be reconstructed from the knowledge of \calTn.

\section{Bond color evolution at boundaries}
\label{sec_colorevolbound}

The preceding discussion makes clear that for 
a system with boundaries 
the formulation of 
Theorem 1 needs to be modified. 
We present and prove these modified theorems below.
We do not attempt at full generality 
but consider only semi-infinite systems with a single
straight boundary, as obtained from figure \ref{fig_horbound} by setting
either $L_+=\infty$ or $L_-=\infty$.\\

\noindent {\bf \sc Theorem \theexerc.} 
\setcounter{refr}{\value{exerc}}
\stepex
{\it Let a triangular lattice \calTzero\ with a receding boundary
have arbitrary positive or negative initial couplings $K_i^{(0)}(\vecr)$.
Then under star-triangle evolution, 
with {\rm (\ref{STTCS})-(\ref{defF})} replaced at the boundary by 
{\rm (\ref{SCtoXY})}, 
the couplings $S_i^{(n)}(\vecr)$ and $C_i^{(n)}(\vecr)$
of the lattices \calTn\ and \calHn, where $n=0,1,2,\ldots$,
remain confined to the union of the real and imaginary axis.}\\

\noindent {\bf \sc Theorem \theexerc.} 
\setcounter{refu}{\value{exerc}}
\stepex
{\it Let a triangular lattice \calTzero\ with an advancing boundary
have arbitrary positive or negative initial couplings $K_i^{(0)}(\vecr)$. 
Then under star-triangle evolution, 
with {\rm (\ref{STTCS})-(\ref{defF})} replaced at the boundary 
by {\rm (\ref{symmsolstan})} or by {\rm (\ref{SCtwothree})} 
and for the random protocol discussed in section 
{\rm \ref{sec_irrevprotocol}},
the couplings $S_i^{(n)}(\vecr)$ and $C_i^{(n)}(\vecr)$ 
of the lattices \calTn\ and \calHn, where $n=0,1,2,\ldots$,
remain confined to the union of the real and imaginary axis.}\\

To prepare for the proof of these two theorems,
we need to slightly extend  Definitions \therefj a and \therefj b\ 
of the \EG\ property in order for these to include the case of a
lattice with boundaries.\\

\noindent {\bf \sc Definition \theexerc }a.  
\setcounter{refs}{\value{exerc}}
{\it A triangular lattice with a receding and/or advancing boundary
is said to be \EGup\ \parEGdown\ if all its 
{\em complete} up-triangles 
({\em complete} down-triangles) are \EGup\ \parEGdown.
The lattice is said to be \EG\ if it is both \EGup\ and \EGdown.}\\

\noindent {\bf \sc Definition \theexerc }b. 
{\it A two-legged star is said to be \EG\ if it contains an even number
(zero or two) of green legs. A hexagonal lattice with a receding and/or
advancing boundary is said to be \EGup\ 
\parEGdown\ if all its complete 
{\em and incomplete} up-stars (down-stars) are \EG.
The lattice is said to be \EG\ if it is both \EGup\ and \EGdown.}\\

It may be noted that Definition \therefs a imposes no condition on
incomplete triangles, but that Definition \therefs b does include a condition 
on incomplete stars.
Definitions \therefp a and \therefp b of the ER property
remain unchanged.
Concerning Definition \therefp a we note 
that at a boundary site at most
four triangle sides may join;
and concerning Definition \therefp b we note 
that all elementary loops are six-legged.\\



\begin{figure}[t]
\begin{center}
\scalebox{.35}
{\includegraphics{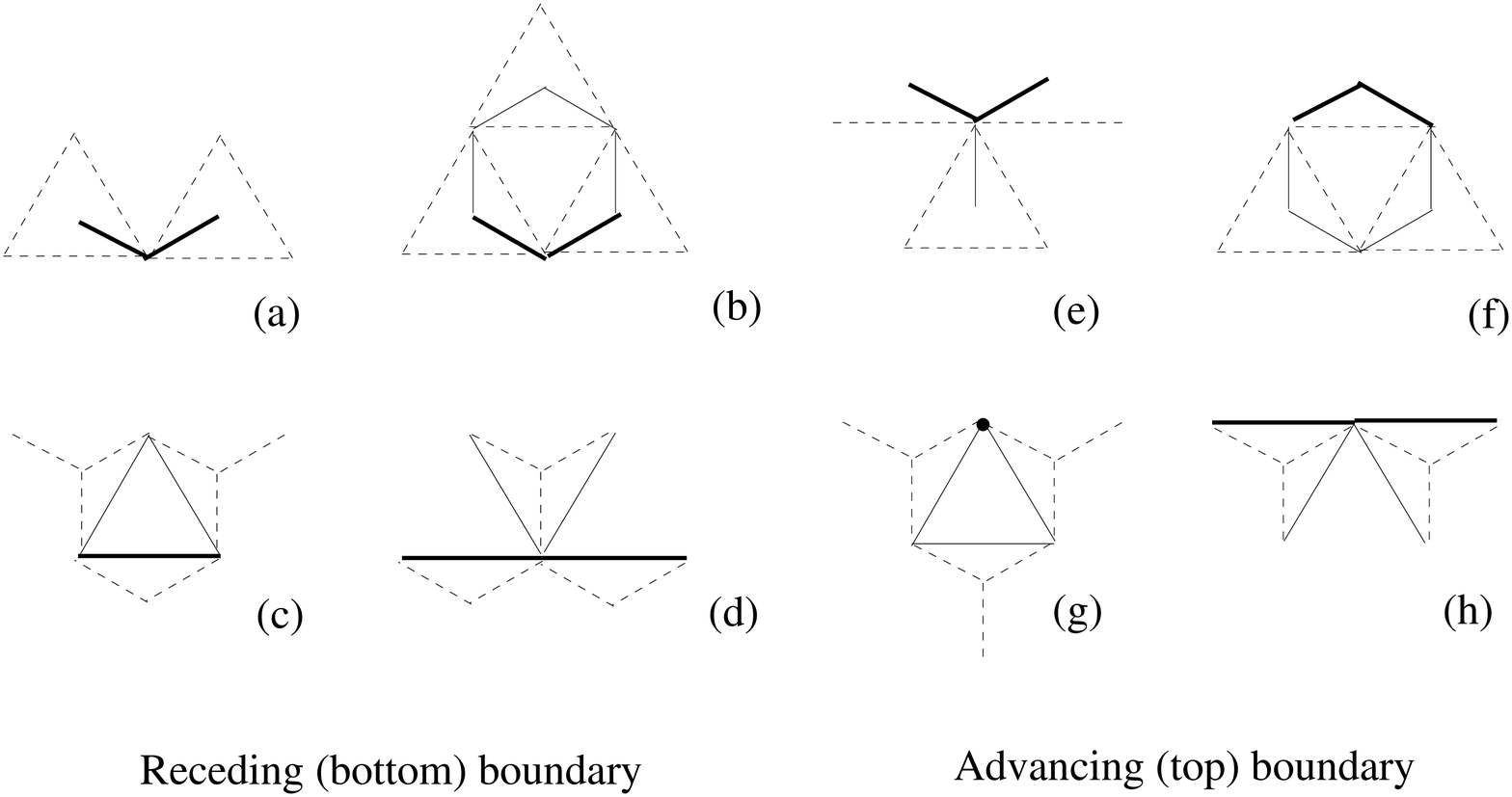}}
\end{center}
\caption{\small
In the proofs of Theorems \therefr\ and \therefu\
we consider figures (a)-(d) and (e)-(h), respectively.
The top row of figures represent the step \calTn\ $\mapsto$ \calHn\
and the bottom row the step \calHn\ $\mapsto$ \calTnpone.
In each figure the solid lines represent the putative
offending configuration;
heaviness of a solid line, and of the dot in figure (g),
indicates that it is located on the boundary.
The dashed lines represent the stars or triangles, complete or incomplete,
that in the preceding step gave rise to the offending configuration. 
The offending configurations are the following:
(a) an incomplete two-legged down-star having exactly one green leg;
(b) a hexagonal loop having an odd number of red legs;
(c) an up-triangle having an odd number of green sides;
(d) a site where four triangle sides join, out of which an odd number is red; 
(e) a down-star having exactly one green leg;
(f) a hexagonal loop having an odd number of red legs;
(g) an up-triangle having an odd number of green sides;
(h) a site where four triangle sides join, out of which an odd number is red.
In each of these eight figures several different colorings are possible.
}
\label{fig_illustr}
\end{figure}

In view of the preceding observations and extended definitions
Theorems \therefr\ and \therefu\ will be proved if we can show that 
in each of the steps \calTn$\,\mapsto$\calHn\ and \calHn$\,\mapsto$\calTnpone,
where $n=0,1,2,\ldots$,
the property of the lattice being EG is transmitted from the ingoing to the
outgoing lattice.  As before, we will show that in fact ER and EG,
now obeying Definitions \therefp\ and \therefs,
respectively, are transmitted together.
Things amount, therefore, to proving the induction step:
if \calTn\ is EG and ER, then so is \calHn, and if \calHn\ is \EG\ and \ER,
then so is \calTnpone.
We prove this for the receding boundary (Theorem \therefr) in section
\ref{sec_receding} and for the advancing boundary (Theorem \therefu)
in section \ref{sec_advancing}.
The reasoning will be along the same lines
as that in section \ref{sec_theoremten},
with many color combinations to be checked separately.
We will illustrate these by the diagrammatic representation
of figure \ref{fig_illustr}, 
which serves the same purpose as 
figures \ref{fig_proofTH} through \ref{fig_proofHTbis}, but is more compact.

\subsection{Receding boundary}
\label{sec_receding}

We consider the semi-infinite lattice obtained by letting $L_+\to\infty$
in figure \ref{fig_horbound}. The boundary at the bottom of this lattice is
receding.\\ 

\noindent {\sc Proof of Theorem \therefr.}

{\it Step 1:} \TtoH. --- Let 
\calTn\ be both EG and ER and suppose that \calHn\ is not.
We will show that this leads to a contradiction. 

Suppose first that \calHn\ is not \EG.
\label{R-HT1}
Since in this step all ingoing up-triangles are complete, 
\calHn\ is necessarily \EGup, and so
we must then have that \calHn\ is not \EGdown.
This means \calHn\ has an offending down-star with an odd number of green 
bonds. If that down-star is in the bulk
we have seen in section \ref{sec_theoremten} 
that this leads to a contradiction,
as was illustrated by figure \ref{fig_proofTH}.
Now let the offending
down-star be an incomplete two-legged star on the bottom boundary of \calHn.
The star legs are shown in figure \ref{fig_illustr}a as solid lines,
their thickness indicating that they are part of the boundary.
The offense of this incomplete star is to have exactly one green leg
(it could be either of the two shown and the figure does not distinguish
between the two possibilities). Now we trace back the origin of these
two legs: in the preceding step they were generated independently
by two distinct up-triangles, shown in the same figure as thin dashed lines.
Upon applying in both triangles Property \therefe\ 
(figure \ref{fig_prop45}a, rotated by $\pm 2\pi/3$ and if necessary 
first mirrored with respect to its central axis)
we see that out of the four triangle sides joining at the center site, 
the number of red ones is either one or three.
This contradicts the premise of \calTn\ being \ER.
It follows that \calHn\ is \EG.

Suppose, secondly, that \calHn\ is not \ER.
\label{R-HT2}
Then there is an elementary six-legged loop 
containing an odd number of red legs.
We have to check the case, shown in figure \ref{fig_illustr}b,
where this loop contains one site of the bottom boundary.
It is easily seen that this leads to a contradiction in the same way as in 
the bulk, where it was illustrated by figure \ref{fig_proofTHbis}.
It follows that \calHn\ is \ER.

Hence \calHn\ is \EG\ and \ER.\\

{\it Step 2:} \HtoT. --- 
Let \calHn\ be both \EG\ and \ER\ and suppose that \calTnpone\ is not. 
We will show that this leads to a contradiction.

Suppose first that \calTnpone\ is not \EG.
\label{R-TH1}
Since this lattice was constructed by ST transformations in down-triangles,
it is necessarily \EGdown.
So we must then have that \calTnpone\ is not \EGup.
This means that \calTnpone\
has an offending up-triangle having either a single or three green
sides, as was illustrated in figure \ref{fig_proofHT} for the bulk.
Suppose now that the offending up-triangle has one side on the bottom
boundary of \calTnpone, as shown in figure \ref{fig_illustr}c. 
This side was generated by an incomplete 
down-triangle, shown by thin dashed lines in the same figure,
and this ST transformation satisfies Property \therefg: 
it contributes a green triangle side if and
only if exactly one of its two legs is red.
Therefore, just as in the bulk, 
we conclude that the hexagonal loop in \calHn\ that surrounds
the offending up-triangle has an odd number of red legs,
which contradicts the premise of \calHn\ being \ER.
It follows that \calTnpone\ is \EG.

Suppose, secondly, that \calTnpone\ is not \ER.
\label{R-TH2}
Then there is an offending site where an odd number of red triangle
sides come together. 
If that site is in the bulk, 
we have seen that this leads to a contradiction, 
as was illustrated by figure \ref{fig_proofHTbis}.
Now let that site be on the boundary, as shown in figure \ref{fig_illustr}d.
It is then connected to four triangle sides out of which
either one or three are red. 
The two horizontal ones of these sides were generated by incomplete
down-stars of \calHn, and the other two by a single complete down-star of
\calHn. 
By using again Property \therefg\ we find that the up-star
of \calHn\ centered at the offending site must have had an odd number of green
legs, which contradicts the premise of \calHn\ being \EG.
It follows that \calTnpone\ is ER.

Hence \calTnpone\ is \EG\ and \ER. This proves Theorem \therefr.
$\square$

\subsection{Advancing boundary}
\label{sec_advancing}

We consider the semi-infinite lattice obtained by letting $L_-\to\infty$
in figure \ref{fig_horbound}. The boundary at the top of this lattice is
advancing.\\

\noindent {\sc Proof of Theorem \therefu.}

{\it Step 1:} \TtoH. -- Let \calTn\ be both EG and ER 
and suppose that \calHn\ is not.
We will show that this leads to a contradiction. 

Suppose first that \calHn\ is not \EG.
\label{A-TH1}
Those ingoing up-triangles of \calTn\
that are complete all lead to up-stars that are \EGup. 
The up-stars of \calHn\ located in the bulk were all generated by complete 
up-triangles of \calTn\ and therefore are \EGup. 
The top boundary of \calHn\ is a zigzagging sequence of star legs;
each pair of legs was generated by a single triangle side
according to the rules symbolized in table \ref{fig_listspec}a,
which show that a pair so generated contains either zero or two green legs.
Each such pair constitutes an incomplete two-legged up-star,
and therefore by Definition \therefs b these incomplete up-stars 
are also \EGup. Therefore, by the same definition, lattice \calHn\ is \EGup.
Since we supposed that \calHn\ is not \EG, it must then fail to be \EGdown.
This means \calHn\ has an offending down-star with an odd number of green bonds.
If that down-star is in the bulk we have seen that this leads to a
contradiction, as was illustrated by figure \ref{fig_proofTH}.
Now let the offending down-star have its two upper legs on 
the top boundary of \calHn, as shown in figure \ref{fig_illustr}e.
These upper legs were generated by two isolated triangle sides,
both shown as thin dashed horizontal lines;
whereas the vertical leg was generated by a full triangle, 
also shown thin dashed.
We may apply Property \therefe\ (see figure \ref{fig_prop45}a) 
to the vertical leg and conclude that if and only if
it is green, then 
it will contribute exactly one red triangle side joining the central site.
And for each of the two upper legs,
we may use table \ref{fig_listspec}a to conclude that if it is green,
then it must have resulted from a triangular side that was red.
It follows that in \calTn\ the down-star center, viewed as a site 
of \calTn, was joined by an odd number of red triangle sides.
This contradicts the premise of \calTn\ being \ER. 
It follows that \calHn\ is \EG.

Suppose, secondly, that \calHn\ is not \ER.
\label{A-TH2}
This means \calHn\ has an offending elementary six-legged loop 
containing an odd number of red legs.
We have seen that if the offending loop is in the bulk, this leads
to a contradiction as illustrated by figure \ref{fig_proofTHbis}.
If the loop is in the top row of loops, as shown in figure \ref{fig_illustr}f,
then the top one of these triangles is single-sided;
nevertheless, together with the two complete up-triangles it encloses a
down-triangle. 
But Property \therefg\ governs in the same way the relation between the red
star legs in the loop and the green triangle sides
in the down-triangle. We find that its number of green sides must be odd,
which contradicts the premise that \calTn\ is \EG.
This leads to a contradiction in the same way as in the bulk.
It follows that \calHn\ is \ER.

Hence \calHn\ is \EG\ and \ER.\\

{\it Step 2:} \HtoT. --- Let \calHn\ be both \EG\ and \ER\ and suppose that
\calTnpone\ is not. We will show that this leads to a contradiction.

Suppose first that \calTnpone\ is not \EG.
\label{A-HT1}
Since this lattice was constructed by ST transformations in down-triangles,
it is necessarily \EGdown.
We must then have that \calTnpone\ is not \EGup.
This means there is
an offending up-triangle having 
either a single or three green sides.
If that up-triangle is in the bulk, this leads to a contradiction,
as illustrated by figure \ref{fig_proofHT}.
Suppose now that the offending up-triangle is in the top row of
up-triangles, as shown in figure \ref{fig_illustr}g,
so that it has one site on the upper boundary. 
Nothing changes and the bulk reasoning applies.
Therefore we conclude that the hexagonal loop surrounding
the offending up-triangle has an odd number of red legs,
which contradicts the premise of \calHn\ being \ER.
It follows that \calTnpone\ is \EG.

Suppose, secondly, that \calTnpone\ is not \ER.
\label{A-HT2}
Then \calTnpone\ must have an offending site 
where an odd number of red triangle sides join. 
If that site is in the bulk, 
we have seen that this leads to a contradiction, 
as illustrated by figure \ref{fig_proofHTbis}.
Now let that site be on the boundary as in figure \ref{fig_illustr}h.
It is then connected to neighboring lattice sites by four bonds, out of which
either one or three are red. 
These sides were generated by two down-stars of \calHn\
shown in the same figure by thin dashed lines.
By using again Property \therefe\ we find that 
out of the two star legs joining at the offending site,
exactly one must be green.
This contradicts the premise of \calHn\ being \EGup.
It follows that \calTnpone\ is \ER.

Hence \calTnpone\ is \EG\ and \ER. 
This proves Theorem \therefu. 
$\square$

\subsection{Green bonds}
\label{sec_greenbonds}

Corollaries \therefk a and \therefk b apply to hexagonal
and triangular lattices, respectively, without boundaries.
In the presence of a receding and/or
an advancing boundary, these Corollaries
have to be modified in the following way.\\

\stepex
\noindent {\sc Corollary \theexerc }a.
\setcounter{refv}{\value{exerc}}
{\it If a hexagonal lattice with a receding and/or advancing boundary
is \EG\ under {\em Definition \therefs b,} then its green bonds constitute 
a collection of loops, and inversely. 
These loops are necessarily
non-self-intersecting and not mutually intersecting}.\\

\noindent {\sc Proof of Corollary \theexerc }a.
The proof is identical to the proof of Corollary \therefk a,
except that the term ``star'' now includes all incomplete two-legged stars.
$\square$\\

\noindent {\sc Corollary \theexerc }b.
{\it If a triangular lattice with a receding and/or advancing boundary
is \EG\ under {\em Definition \therefs a}, then the duals of its green bonds 
constitute a collection of loops and chains on its dual hexagonal\,}%
\footnote{See footnote \ref{footnote1}.}
{\it lattice, the chain ends being in a one-to-one correspondence with the
green boundary bonds; and inversely.
These loops and chains are necessarily non-self-intersecting and not
mutually intersecting.}\\

\noindent {\sc Proof of Corollary \theexerc }b.
Assign to each bond of the dual hexagonal 
lattice the color of the corresponding primal bond.
This dual lattice thus colored is hexagonal and 
each of its full stars, since it corresponds to an \EG\ triangle, 
is therefore \EG\ itself. 
However, the lattice has vertical star legs sticking out of the boundary
and that belong to a complete up-star {\it or\,} to a complete down-star,
but not both. The proof is as that of Corollary \therefk b,
except that when one of the legs in the sequence 
$\alpha, \beta, \gamma, \ldots$ is such a vertical leg, then the 
construction stops and we have reached the end of the chain.
$\square$

\subsection{Parallel and intersecting boundaries}
\label{sec_slabs}

In order to extend in
Sections \ref{sec_receding} and \ref{sec_advancing}
the proof of Theorem \therefa\ to a system with a boundary,
we only had to consider special bond configurations 
located at that boundary.
In the presence of two parallel boundaries
the same arguments hold independently for each of the two boundaries.
We may therfore conclude that also in a slab the bond variables
$(S_i^{(0)}(\vecr),C_i^{(0)}(\vecr))$ remain confined to the union of the 
real and the imaginary axis. 
We will not spell this out in detail.

A single straight boundary at an angle of $\pi/3$ or $2\pi/3$ 
with respect to the horizontal axis 
may be reduced to one of the two above boundary types
by an appropriate rotation.
If two intersecting boundaries are present,
the system fills a wedge (or its complement),
and the ST transformation rule for stars and triangles located
at the tip of the wedge has to be considered.
We will not consider here such more general configurations of boundaries.
We believe that in all cases it is possible to extend
our main theorem and prove that the $\{S_i^{(n)}(\vecr),C_i^{(n)}(\vecr)\}$
remain confined to the union of the real and the imaginary axis.
A specific example is the lattice 
studied (but restricted to ferromagnetic interactions)
in references \cite{HSvL78,HSvL79,KH79,YH79}, 
which fills a finite triangular domain delimited by three
intersecting receding boundaries.
For the disordered version of this lattice the appropriate
extension of Theorem \therefa\ are readily proven.

\begin{figure}[ht]
\begin{center}
\scalebox{.50}
{\includegraphics{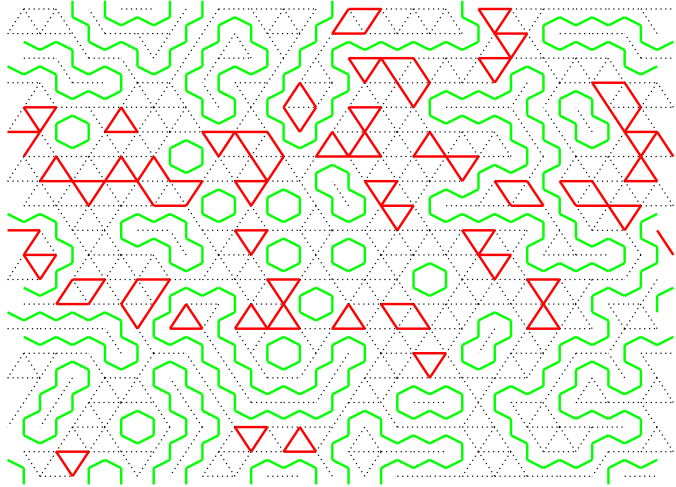}}
\end{center}
\caption{\small
Snapshot of the bond coloring of lattice $\calT(10)$.
The lattice is a slab with a periodicity of $20$ lattice units
in the horizontal direction
and of finite width $L=20$ in the vertical direction.
For greater clarity, instead of drawing the green bonds of $\calT(10)$
themselves, we have drawn their duals.
Most of the green chains form loops, but some are seen to be open-ended at the
boundaries, in agreement with Corollary \therefv b.
Each site of $\calT(10)$ is joined by an even number, possibly zero, 
of red bonds.
See text for more details.
}
\label{fig_snapbound}
\end{figure}

\subsection{Numerical evaluation}
\label{sec_numericalbound}

We have evaluated the ST evolution numerically for the slab depicted in 
figure \ref{fig_snapbound}. 
The top boundary advances and the bottom boundary recedes,
the width $L$ remaining constant.
The bond strengths $S_i^{(0)}(\vecr)$ of the initial lattice \calTzero\ 
were drawn independently from a zero-mean Gaussian distribution $N(S_i)$
of width $\langle S_i^2\rangle^{1/2}=2$, so that all initial bonds were black. 
Figure \ref{fig_snapbound} shows the bond colors of lattice
$\calT(10)$ that result after ten iterations with the uniform random protocol
of section \ref{sec_irrevprotocol} applied at the top boundary.
The green star legs form chains: a part of them are closed loops and another
part begin and end at the top and/or bottom boundaries,
in agreement with Corollary \therefv b.

\section{Conclusion}
\label{sec_conclusion}

The iterated star-triangle (ST) transformation  
is a tool that has led to various interesting results on 2D Ising systems.
The transformation may be applied to triangular and hexagonal lattices 
with arbitrarily inhomogeneous nearest-neighbor couplings $K_i^{(0)}(\vecr)$. 
To the best of our knowledge, however, there have not been, as of today, 
any applications to frustrated lattices, 
the difficulty being that in such lattices
iteration may lead to complex-valued couplings.
In this work we have addressed this problem.

We have found a general theorem, together with several corollaries,
that is obeyed by systems that are subjected to iterated ST transformation
(called in this work ``ST evolution'').
Our main result is that 
under ST evolution, when applied to an initial system with real-valued 
bond strengths $K_i^{(0)}(\vecr)$, 
the variables $S_i^{(n)}(\vecr)=1/\sinh 2K_i^{(0)}(\vecr)$ that are 
generated successively will remain
confined to the union of the real and the imaginary axis.
This property is of interest in itself but it is also
important for numerical applications.
The result was shown to hold for systems which are infinite or 
have periodic boundary conditions, and also, when suitably reformulated,  
for half-infinite systems. 
In the latter the boundary behavior under ST evolution requires special 
attention and occupies a large place in this work.
A distinction appears between receding and advancing boundaries:
advancing ones require a protocol prescribing how new
bonds at the boundary are created.
We have shown that there is a class of random protocols that ensures 
the validity of the main result in the presence of an advancing boundary.
Our method of proof attributes
to each bond one out of three  ``colors,'' 
black, red, or green, according to the value of its bond strength.
Under ST evolution these colors obey rules
that are instrumental in proving the theorems.
One might think that our results ought to follow from a short and
simple symmetry argument. If one exists, however, we think it would be
unlikely to automatically cover the boundary behavior.

We illustrated our work by two brief numerical calculations, 
one for a lattice with periodic boundary conditions in both directions, 
and another one for a slab, periodic in only one direction 
and obeying a specific ``uniform random'' protocol 
along its advancing boundary.\\

We list a few interesting questions have not been answered.
First, Section \ref{sec_irrevprotocol} shows that at an advancing
boundary we have much freedom to define a protocol. 
In Section \ref{sec_numericalbound}
we used the uniform random one, which is the first one to come to mind;
but, can anything be gained from other protocols?
And, well beyond the scope of this paper,
would there be any advantage in employing
protocols that do {\it not\,} confine the $S_i^{(n)}(\vecr)$ 
to the union \un of the two main axes in the complex plane?
Secondly, there is the open question, mentioned in Section
\ref{sec_numerical}, of how to extract, with the aid of the present
techniques, the zero-temperature exponents of the 2D Ising spin glass.
Thirdly, there is a factor $\exp(g)$ that multiplies the
partition function each time an ST transformation is applied.
The study and the exploitation of this factor have been left aside in the
present work, but certainly deserve attention;
it enters, in particular, the calculation of free energies.
Fourth, there are many quantitative numerical questions. 
How does an initial distribution of coupling strengths evolve?  
What is their asymptotic distribution, if any,
when the number $n$ of iterations tends to infinity, 
and how does this depend on whether or not the lattice has a boundary?

We believe that the answer to several of these and to other questions,
as well as specific applications, 
constitute an interesting direction of future research.




\appendix



\begin{thebibliography}{10}

\bibitem{Syozi72}
I. Syozi, 
in {\it Phase Transitions and Critical Phenomena},
edited by C. Domb and M.S. Green (Academic, London, 1972), Vol.1, p.
269.

\bibitem{Onsager44}
L.~Onsager,
{\it Phys. Rev.} {\bf 65} (1944) 117.

\bibitem{Wannier45}
G.H.~Wannier,
{\it Rev. Mod. Phys.} {\bf 17} (1945) 50.

\bibitem{Houtappel50}
R.M.F.~Houtappel, 
{\it Physica\,} {\bf 16} (1950) 425. 

\bibitem{AuYangPerk89}
H.~Au-Yang and J.H.H.~Perk,
{\it Advanced Studies in Pure Mathematics\,} {\bf 19} (1989) 57.

\bibitem{PerkAuYang06}
J.H.H.~Perk and H.~Au-Yang,
{\it Encyclopedia of Mathematical Physics,} eds. J.-P.-Fran\c{c}oise, 
G.L.~Naber and S.T.~Tsou, Oxford: Elsevier, 2006 
(ISBN 978-0-1251-2666-3), volume 5, pages 465-473.

\bibitem{HSvL78} 
H.J.~Hilhorst, M. Schick, and J.M.J. van Leeuwen,
{\it Phys. Rev. Lett.} {\bf 40} (1978) 1605.

\bibitem{HSvL79} 
H.J.~Hilhorst, M. Schick, and J.M.J. van Leeuwen,
{\it Phys. Rev. B\,} {\bf 419} (1979) 2749.

\bibitem{KH79} 
H.J.F.~Knops and H.J.~Hilhorst, 
{\it Phys.Rev. B} {\bf 19} (1979) 3689.

\bibitem{YH79} 
Y.~Yamazaki and H.J.~Hilhorst, 
{\it Phys. Lett.} {\bf 70A} (1979) 329.

\bibitem{YMH79} 
Y.~Yamazaki, G.~Meissner, and H.J.~Hilhorst,  
{\it Z. Phys.} {\bf B 35} (1979) 333.

\bibitem{YMH80} 
Y.~Yamazaki, H.J.~Hilhorst, and G.~Meissner, 
{\it J. Stat. Phys.} {\bf 23} (1980) 609.

\bibitem{HvL81}
H.J.~Hilhorst and J.M.J.~van~Leeuwen,
{\it Phys. Rev. Lett.} {\bf 47} (1981) 1188.

\bibitem{Burkhardt82}
T.W.~Burkhardt,
{\it Phys. Rev. Lett.} {\bf 48} (1982) 216.

\bibitem{Cordery82}
R.~Cordery,
{\it Phys. Rev. Lett.} {\bf 48} (1982) 215.

\bibitem{BGHvL84} 
T.W.~Burkhardt, I.~Guim, H.J.~Hilhorst, and J.M.J.~van Leeuwen, 
{\it Phys.Rev. B} {\bf 30} (1984) 1486.

\bibitem{BurkhardtGuim84}
T.W.~Burkhardt and I.~Guim,
{\it Phys. Rev. B\,} {\bf 29} (1984) 508(R).

\bibitem{BurkhardtGuim85}
T.W.~Burkhardt and I.~Guim,
\newblock {\it J. Phys. A\,} {\bf 18} (1985) L33.

\bibitem{IgloiLajko96}
F.~Igl\'oi and P.~Lajk\'o,
{\it J. Phys. A\,} {\bf 29} (1996) 4803.

\bibitem{BurkhardtGuim98}
T.W. Burkhardt and I. Guim, 
\newblock {\it Physica A\,} {\bf 251} (1998) 12.

\bibitem{LajkoIgloi00}
P.~Lajk\'o and F.~Igl\'oi,
{\it Phys. Rev. E\,} {\bf 61} (2000) 147. 

\bibitem{BercheTurban90}
B.~Berche and L.~Turban,
{\it J. Phys. A\,} {\bf 23} (1990) 3029.

\bibitem{Turbanetal94}
L.~Turban, F.~Igl\'oi, and B.~Berche,
{\it Phys. Rev. B\,} {\bf 49} (1994) 12695.

\bibitem{Selkeetal97}
W.~Selke, F.~Szalma, P.~Lajk\'o, , and F.~Igl\'oi,
{\it J. Stat. Phys.} {\bf 89} (1997) 1079.

\bibitem{Igloietal98}
F.~Igl\'oi, P.~Lajk\'o, W.~Selke, and F.~Szalma,
{\it J. Phys. A\,} {\bf 31} (1998) 2801.

\bibitem{Turbanetal99}
L.~Turban, D.~Karevski, and F.~Igl\'oi,
{\it J. Phys. A\,} {\bf 32} (1999) 3907.

\bibitem{Karevskietal00}
D.~Karevski, L.~Turban, and F.~Igl\'oi,
{\it J. Phys. A\,} {\bf 33} (2000) 2663.

\bibitem{Turban02}
L.~Turban,
{\it Physica A\,} {\bf 303} (2002) 275.

\bibitem{Colluraetal09}
M.~Collura, D.~Karevski, and L.~Turban,
{\it J. Stat. Mech.} (2009) P08007.

\bibitem{Igloietal93}
F.~Igl\'oi, I.~Peschel, and L.~Turban,
{\it Advances in Physics\,} {\bf 42} (1993) 683.
Taylor \& Francis.

\bibitem{Pelizzola97}
A.~Pelizzola,
{\it Int. J. Mod. Phys. B\,} {\bf 11} (1997) 1363.

\bibitem{FisherHuse86}
D.S.~Fisher and D.A.~Huse,
{\it Phys. Rev. Lett.} {\bf 56} (1986) 1601.

\bibitem{FisherHuse88}
D.S.~Fisher and D.A.~Huse,
{\it Phys. Rev. B\,} {\bf 38} (1988) 386.

\bibitem{ThillHilhorst96}
M.J.~Thill and H.J.~Hilhorst,
{\it J. Phys. I France} {\bf 6} (1996) 67.

\end{thebibliography}
\end{document}